\documentclass[%
 reprint,
 amsmath,amssymb,
 aps,
 pre,
showkeys
]{revtex4-2}

\usepackage{graphicx}
\usepackage{bm}
\usepackage{xcolor}
\usepackage{mathtools}
\usepackage{xcolor}
\usepackage[colorlinks=true,linkcolor=blue,citecolor=blue,urlcolor=blue]{hyperref}

\usepackage[mathlines]{lineno}

\begin{document}


\title{Effect of higher-order interactions on noisy majority-rule dynamics with random group sizes}

\author{Roni Muslim\textsuperscript{1,2}}
\email{contact: roni.muslim@apctp.org}
\author{Jong-Min Park\textsuperscript{1,3}}
\email{contact: jongmin.park@apctp.org}
\author{Jihye Kim\textsuperscript{4}}
\email{contact: arrr3755@naver.com}
\author{Rinto Anugraha NQZ\textsuperscript{5}}
\email{contact: rinto@ugm.ac.id}

\affiliation{\textsuperscript{1}Asia Pacific Center for Theoretical Physics,  Pohang, 37673, Republic of Korea}
\affiliation{\textsuperscript{2} Research Center for Quantum Physics, BRIN, South Tangerang, 15314, Indonesia}
\affiliation{\textsuperscript{3} Department of Physics, POSTECH, Pohang 37673, Republic of Korea}
\affiliation{\textsuperscript{4}Department of Physics, Korea University, Seoul, 02841, Republic of Korea}
\affiliation{\textsuperscript{5}Department of Physics, Universitas Gadjah Mada, Yogyakarta, 55282, Indonesia}

\date{\today}

\begin{abstract}
We study opinion dynamics with higher-order interactions, motivated by the fact that social influence often takes place in groups rather than only through pairwise contacts. We introduce a noisy majority-rule model on annealed hypergraphs with heterogeneous group sizes and investigate how the distribution of interaction sizes affects collective ordering and relaxation. Using analytical theory and Monte Carlo simulations, we show that group-size heterogeneity strongly shapes both the transition between ordered and mixed states and the associated time scales. In particular, broader and heavier-tailed distributions make ordering more robust by enhancing the effect of rare large-group events. They also modify the finite-size scaling of relaxation, producing a crossover from the standard logarithmic behavior to faster ordering in sufficiently broad ensembles. In the pure majority-rule limit, we further show that the exit probability near coexistence obeys a universal error-function scaling form controlled by a single structural parameter. Our results demonstrate that the full distribution of group sizes is a key determinant of nonequilibrium ordering in higher-order opinion dynamics.
\end{abstract}

\maketitle

\section{\label{sec:intro} Introduction}
Binary-state opinion models provide a minimal yet powerful framework for probing nonequilibrium collective phenomena in interacting-agent systems~\cite{castellano2009statistical,liggett2013stochastic,galam2012socio}. Among them, majority-rule (MR) dynamics, also known as the Galam majority rule, serves as a canonical baseline for consensus formation~\cite{galam1986majority,galam2008sociophysics,cheon2018dynamical,krapivsky2003dynamics,chen2005majority}. In this model, a set of agents simultaneously adopts its local majority opinion at each update. The rule captures the tension between deterministic alignment induced by local majorities and finite-size stochasticity due to random sampling, thereby shaping macroscopic ordering and relaxation~\cite{lambiotte2007majority,nguyen2020dynamics,mulya2024phase,muslim2024impact,azhari2023external}. In well-mixed populations, MR yields transparent, analytically tractable predictions for collective outcomes and characteristic time scales, making it a standard reference for social alignment processes and a natural point of departure for higher-order generalizations~\cite{crokidakis2015inflexibility,muslim2022phase,oestereich2023phase}.

A salient empirical regularity of social interaction is its polyadic nature: deliberation and decision-making often occur in groups rather than dyads. Hypergraphs and simplicial complexes provide explicit combinatorial representations of such higher-order contacts and have revealed qualitatively new dynamical regimes across contagion and synchronization, including discontinuous social contagion and explosive or abrupt transitions~\cite{battiston2020networks,boccaletti2023structure,bick2023higher,iacopini2019simplicial,de2020social,skardal2020higher,millan2020explosive,zhang2023higher,majhi2022dynamics}. Empirical face-to-face datasets further document group interactions and heterogeneous meeting sizes, underscoring the need for higher-order formalisms~\cite{stehle2011high,mastrandrea2015contact,patania2017shape,benson2016higher,benson2018simplicial,roh2023growing}.
Mechanistically related conformity--field and conformity--noise update rules have also been used in agent-based models of collective social dynamics and innovation adoption~\cite{byrka2016difficulty,abramiuk2023rigorous}, albeit in a different application context from the noisy hypergraph MR dynamics studied here. In the MR setting, hypergraph formulations make it possible to connect microscopic group updates with mesoscopic drift--diffusion descriptions; for fixed group sizes, such approaches reproduce simulations with high fidelity and clarify how higher-order interactions shape ordering pathways~\cite{noonan2021dynamics,lanchier2013stochastic,neuhauser2022consensus,de2020social}. These developments motivate extending the framework to heterogeneous group-size distributions.

The same framework also admits a concrete social reading. Hyperedges represent recurrent arenas of discussion, such as families, committees, work teams, or group-messaging threads, while the hyperedge-size distribution $P(n)$ encodes heterogeneity in forum sizes. An ``annealed'' representation captures rapidly reconfigured contexts (e.g., rotating committees or fluid online threads), whereas a ``quenched'' representation reflects persistent group memberships; both are standard coarse-graining choices in nonequilibrium dynamics on networks and higher-order structures~\cite{pastor2015epidemic,gleeson2013binary,battiston2020networks,neuhauser2022consensus}. In either view, sampling rules and the prevalence of particular group sizes modulate exposure to majority cues, the stability of local agreements, and the pathways to population-level consensus, in line with established insights on group-size effects, classic conformity experiments, and higher-order social contagion~\cite{asch2016effects,mercier2019majority,iacopini2019simplicial,noonan2021dynamics,hartnett2016heterogeneous}.

Randomness in decision making constitutes a second key ingredient. In the majority-vote (MV) family, a noise parameter competes with local alignment and drives order--disorder transitions whose thresholds and critical properties depend on topology and heterogeneity~\cite{de1992isotropic,chen2015critical,chen2020non}. Socially, such randomness can represent private information, limited attention, exogenous messaging, or institutional guidance that biases updates away from the locally preferred opinion. Although MV provides a canonical benchmark for noise-driven symmetry breaking in opinion dynamics, the coupling of such biased external-field updates to higher-order, majority-driven group dynamics on hypergraphs remains comparatively underexplored, especially beyond fixed group sizes and within frameworks that admit drift--diffusion closures~\cite{crokidakis2015inflexibility,muslim2022phase,oestereich2023phase,battiston2020networks,boccaletti2023structure,noonan2021dynamics}.

Classical results for MR provide the baseline against which noise and higher-order effects can be assessed: on complete graphs, the mean consensus time $T_{\mathrm{cons}}$ grows logarithmically with system size, and the exit probability approaches a step function in the thermodynamic limit~\cite{krapivsky2003dynamics,chen2005majority}. These benchmarks clarify what must change once a biased external field is introduced and group sizes are allowed to vary. What remains lacking, however, is a systematic account of noisy MR on hypergraphs with random hyperedge sizes: how a field-driven group update interacts with an empirically broad $P(n)$ to set the symmetric ordering threshold $q_c(P)$ at $p=1/2$, and, more generally, how it organizes relaxation, bias-selected attractors, and exit probabilities away from the symmetric point in socially heterogeneous environments~\cite{patania2017shape,benson2016higher,benson2018simplicial,roh2023growing,mercier2019majority}.

To close this gap, we develop a unified formulation of noisy majority rule on annealed hypergraphs with random group sizes. At each update, a group size $n$ is drawn from a prescribed distribution $P(n)$, after which $n$ distinct agents are selected uniformly at random from the population without replacement. The selected hyperedge then undergoes either a local majority-rule update or a collective field-driven update controlled by the parameters $q$ and $p$. We retain the explicit $P(n)$ dependence throughout and combine a mean-field (MF) drift--diffusion theory with Monte Carlo (MC) simulations. We focus on the symmetric ordering threshold $q_c(P)$ at $p=1/2$, the consensus and disordering time scales, and the exit probability in the pure majority-rule limit. This framework links microscopic group-size heterogeneity to macroscopic ordering, relaxation, and first-passage behavior in noisy higher-order opinion dynamics.

\section{\label{sec:model} Model}

We consider a population of $N$ agents, each holding a binary opinion $\sigma_i\in\{+1,-1\}$. Interactions occur through hyperedges whose sizes are random. At each elementary update, a group size $n$ is drawn from a prescribed distribution $P(n)$ supported on $\{n_{\min},\dots,n_{\max}\}$, with $n_{\min}\ge 3$ and $n_{\max}\le N$, and $n$ distinct agents are sampled uniformly at random from the population without replacement. We adopt an annealed (well-mixed) representation, in which a fresh hyperedge is resampled at every update. One Monte Carlo step (MCS) consists of $N$ such elementary updates. Let $t=0,1,2,\dots$ denote the number of elementary updates and define the rescaled time $\tau\equiv t/N$, so that one MCS corresponds to $\Delta\tau=1$ and each elementary update advances time by $\delta\tau=1/N$.

We consider three canonical families of $P(n)$: (i) the $n$-uniform case, $P_{\mathrm u}(n')=\delta_{n',n}$, with fixed hyperedge size $n$; (ii) the truncated shifted geometric distribution, $P_{\mathrm{geo}}(n)=(1-x)x^{\,n-3}/(1-x^{\,n_{\max}-2})$ for $n=3,\dots,n_{\max}$ and $0<x<1$, whose thermodynamic-limit mean is $\langle n\rangle=3+x/(1-x)$; and (iii) the truncated power law, $P_{\mathrm{pl}}(n)=n^{-\alpha}/Z_\alpha$, with $Z_\alpha=\sum_{k=n_{\min}}^{n_{\max}} k^{-\alpha}$ and $\alpha>0$, motivated by empirical datasets~\cite{patania2017shape,benson2018simplicial,roh2023growing}. The cutoff $n_{\max}$ enforces feasibility and may be taken as $n_{\max}=N$ or, more generally, as a sublinear cutoff such as $n_{\max}=\lfloor N^\eta\rfloor$ with $0<\eta\le 1$. Throughout the main text we set $n_{\min}=3$ to isolate genuinely higher-order interactions beyond pairwise links. With the strict-majority rule and no update in tie events used here, the pairwise case $n=2$ is dynamically degenerate in the majority-rule branch; hence the smallest nontrivial majority-rule interaction in the present framework occurs at $n=3$. Both odd and even group sizes are allowed; for even $n$, tie events are treated explicitly below.

\begin{figure}[t]
    \centering
    \includegraphics[width=0.55\linewidth]{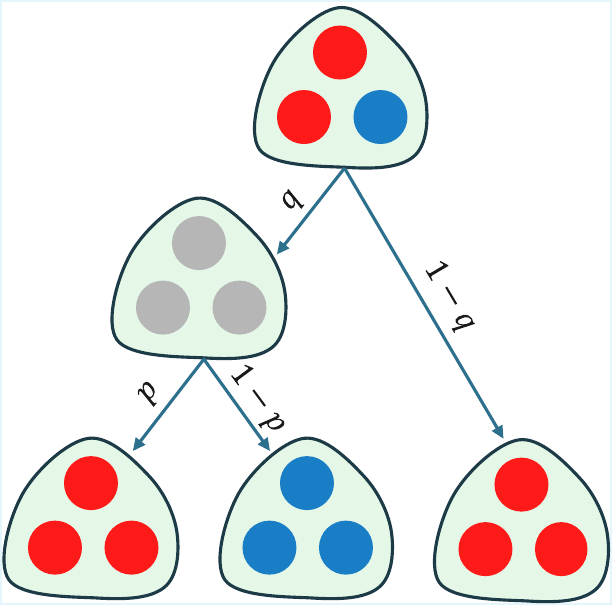}
    \caption{Schematic illustration of a single update (shown here for $n=3$). A group size $n$ is drawn from $P(n)$, after which $n$ distinct agents are sampled uniformly at random without replacement. The selected hyperedge then follows either a collective field-driven update, setting all selected agents to $+1$ with probability $qp$ or to $-1$ with probability $q(1-p)$, or the local majority rule with probability $1-q$. Red and blue nodes denote states $+1$ and $-1$, respectively, while gray nodes indicate the selected agents prior to the update. Ties for even $n$ yield no change (not shown).}
    \label{fig:model_schematic}
\end{figure}

Given the selected hyperedge $\{i_1,\dots,i_n\}$, we define the pre-update sum as $\Sigma=\sum_{j=1}^{n}\sigma_{i_j}$. With probability $q$, the selected group undergoes a field-driven update: all selected agents adopt state $+1$ with probability $p$ and state $-1$ with probability $1-p$. With complementary probability $1-q$, all selected agents adopt the strict majority state of the selected hyperedge: $\sigma_{i_j}=+1$ if $\Sigma>0$, $\sigma_{i_j}=-1$ if $\Sigma<0$, and no change occurs if $\Sigma=0$. For the $n$-uniform case $n=3$, the model reduces to a triplet majority-rule dynamics with a competing field-driven group update~\cite{forgerini2024directed}. A schematic illustration is shown in Fig.~\ref{fig:model_schematic}.

The model is controlled by two parameters: $q\in[0,1]$, the probability of the field-driven branch, and $p\in[0,1]$, the bias toward state $+1$ in that branch. As the main macroscopic observable, we track the fraction $c(\tau)$ of agents in state $+1$. Unless stated otherwise, initial conditions are prepared by assigning state $+1$ independently with probability $c_0$ and state $-1$ with probability $1-c_0$, so that $c_0=1/2$ corresponds to a balanced initial condition in expectation, whereas $c_0\in\{0,1\}$ gives fully ordered initial states. For the exit-probability simulations in Sec.~\ref{sec:exit_prob}, the initial fraction is fixed by preparing $N_+=c_0N$ agents in state $+1$, with $c_0N$ chosen to be an integer, and randomly permuting their positions. Further details of the MC implementation, including the preparation of initial conditions and the stopping criteria used for the measured first-passage observables, are summarized in Appendix~\ref{app:MC_protocol}.

In the limiting cases, $q=0$ recovers pure majority-rule dynamics, whereas $q=1$ yields purely field-driven collective updates with thermodynamic fixed point $c^\ast=p$. The case $p=1/2$ is up-down symmetric, while $p\neq 1/2$ represents a directional field. Throughout this work, the symbol $q$ denotes the probability of the field-driven branch and should not be confused with the lobby size in the standard $q$-voter model~\cite{muslim2025ordering}.

\section{\label{sec:section_iii} Transition probabilities and criticality}

\subsection{\label{sec:sub_section_iii_i} Transition probabilities}

At the microscopic level, the dynamics is described by a master equation for the probability distribution of the discrete variable $k$, the number of agents in state $+1$. Since one elementary hyperedge update may change several agents simultaneously, the process is not a single-spin birth--death process but a multispin jump process. The first-passage quantities studied in this work, such as the consensus time and the disordering time, correspond to the first arrival of $k$ at specific values, such as $k=0$, $k=N$, or the neighborhood of $k=N/2$.

In the annealed setting considered here, the interacting group is resampled uniformly at random from the whole population at every update. Thus, local correlations are continually destroyed and do not build up over time, so the pre-update composition of the active group is determined solely by the instantaneous global density $c=k/N$. This annealed mixing property makes the dynamics effectively MF at the mesoscopic level and explains why the analytical theory is expected to be close to the MC results in most parameter regimes.

For large $N$, it is convenient to introduce the intensive variable $c=k/N$. When the typical change in $k$ produced by one update is small compared with $N$, the corresponding increment $\Delta c$ is also small, and the master equation can be approximated by a Fokker--Planck (FP) equation in $c$-space. This diffusion approximation amounts to truncating the Kramers--Moyal (KM) expansion at second order, or equivalently retaining only the first two conditional jump moments of $\Delta c$~\cite{van1992stochastic,gardiner2009stochastic,risken1996fokker}. The first moment gives the drift, while the second central moment gives the diffusion coefficient describing finite-size fluctuations.

The FP approximation is expected to be most accurate in the annealed MF regime for large $N$ and moderate effective jump sizes. For very broad hyperedge-size distributions, occasional large sampled groups may generate jumps in $c$ for which higher-order jump moments become less negligible. We nevertheless use the FP framework because it provides a compact mesoscopic route to the drift, linear stability analysis, and the backward-FP formulation of first-passage observables such as exit probabilities, consensus times, and disordering times~\cite{gardiner2009stochastic,risken1996fokker}.

Denoting by $N_{-\to +}$ and $N_{+\to -}$ the numbers of $-\!\to +$ and $+\!\to -$ flips produced by one elementary update, the change in $c$ is
$\Delta c = (N_{-\to +}-N_{+\to -})/N$. Using the rescaled MC time $\tau=t/N$ (so that $\delta\tau = 1/N$ per elementary update), the drift and diffusion coefficients in the diffusion approximation are defined by
\begin{align}
  v(c) &= \frac{\langle \Delta c\rangle}{\delta \tau}
        = N\,\langle \Delta c\rangle,  \label{eq:v_def_FP}\\[2pt]
  D(c) &= \frac{\langle (\Delta c)^2\rangle
             - \langle \Delta c\rangle^2}{\delta \tau}
        = N\,\mathrm{Var}[\Delta c],
  \label{eq:D_def_FP}
\end{align}
where $\langle\cdot\rangle$ denotes an average over the hyperedge selection and the dynamical noise, and
$\mathrm{Var}[\Delta c]$
is the variance of the one-step increment $\Delta c$ conditioned on the current state $c$.

Equivalently, if $\mathcal{P}(c,\tau)$ denotes the probability density of the mesoscopic state, then $v(c)$ and $D(c)$ are the first two conditional jump moments entering the effective FP description. Because the active group is redrawn randomly at every step, these conditional moments depend only on the instantaneous global density $c$, not on persistent local configurations.

It is convenient to express these moments in terms of the expected numbers of upward and downward flips per update, defined respectively as
$R(c) \equiv \langle N_{-\to+}\mid c\rangle$ and $L(c) \equiv \langle N_{+\to-}\mid c\rangle$.
Introducing $\Delta N \equiv N_{-\to +}- N_{+ \to -}$, Eqs.~\eqref{eq:v_def_FP} and \eqref{eq:D_def_FP} can be rewritten as
\begin{align}
  v(c) &= R(c)-L(c),  \label{eq:v_from_RL_main}\\[2pt]
  D(c) &= \frac{1}{N} \left[\left\langle(\Delta N)^{2}\right\rangle - \langle \Delta N \rangle^2 \right].
  \label{eq:D_from_RL_exact}
\end{align}

For the without-replacement sampling used here, the exact finite-$N$ composition of a selected hyperedge is hypergeometric. In the annealed large-$N$ MF closure, however, this composition is approximated by the binomial form
$B_{n,\ell}(c)=\binom{n}{\ell}c^\ell(1-c)^{n-\ell}$.
This approximation becomes asymptotically exact in the well-mixed limit when $n/N\to 0$ and provides the analytical closure used below. Finite-$N$ corrections associated with the exact hypergeometric sampling are therefore not retained explicitly in the compact MF expressions.

Within this annealed MF closure, the two update branches contribute as follows. In the majority branch, with probability $1-q$, if $\ell>\lfloor n/2\rfloor$, the number of upward flips is $n-\ell$, whereas if $\ell<n/2$, equivalently $\ell\le \lfloor (n-1)/2\rfloor$, the number of downward flips is $\ell$; ties produce no update. In the field-driven branch, with probability $q$, when the collective $+1$ outcome is chosen, the average upward contribution is proportional to the fraction $1-c$ of agents in state $-1$ in the selected hyperedge, while the collective $-1$ outcome gives the corresponding downward contribution proportional to $c$.

This gives the corresponding closure expressions
\begin{align}
  R(c) &\approx \sum_{n=3}^{n_{\max}} P(n)\Big[\left(1-q\right)M_n(c)
          + qnp\left(1-c\right)\Big], \label{eq:R_defs_main} \\
  L(c) &\approx \sum_{n=3}^{n_{\max}} P(n)\Big[\left(1-q\right)M_n\left(1-c\right)
          + qn\left(1-p\right)c\Big], \label{eq:L_defs_main}
\end{align}
where $  M_n(c)=\sum_{\ell=\lfloor n/2\rfloor+1}^{n}
  (n-\ell)\binom{n}{\ell}\,c^{\ell}(1-c)^{n-\ell}$ is the MF majority-rule contribution associated with a strict majority of $+1$ opinions in a group of size $n$, with ties producing no update, while $M_n(1-c)$ gives the corresponding contribution for a strict majority of $-1$. Within the same closure, the field-driven terms in Eqs.~\eqref{eq:R_defs_main}--\eqref{eq:L_defs_main} reflect that, on average, fractions $1-c$ and $c$ of the selected group are initially in states $-1$ and $+1$, respectively.

Substituting Eqs.~\eqref{eq:R_defs_main}--\eqref{eq:L_defs_main} into Eq.~\eqref{eq:v_from_RL_main} gives the corresponding closure expression for the drift,
\begin{equation}
  v(c)\approx\sum_{n=3}^{n_{\max}} P(n)\left\{ (1-q)\,\Delta M_n(c)
  + q\,n\left(p-c\right) \right\},
  \label{eq:drift_general}
\end{equation}
where $\Delta M_n(c)\equiv M_n(c)-M_n(1-c)$.
The second term in braces is the field-driven contribution, which shifts the dynamics toward the externally preferred value $c=p$ in expectation, whereas the first term represents the ordering tendency of strict majority rule. Both effects are averaged over the hyperedge-size distribution $P(n)$. For $p=1/2$, the field-driven branch restores the symmetric mixed state. For $p\neq 1/2$, it explicitly breaks the up--down symmetry and shifts the dynamics toward the side favored by the external field. At the extreme biases $p=1$ and $p=0$, this field-driven contribution acts maximally toward the corresponding consensus states, whereas for $0<p<1$ it pulls the system toward the interior value $c=p$.

For later use, within the same binomial large-$N$ MF closure, we approximate the second jump moment $\langle (\Delta N)^2\rangle$ by $\mathcal{S}(c)$. After averaging over the group-size distribution, this closure gives
\begin{equation}
  \mathcal{S}(c) \approx \sum_{n=3}^{n_{\max}}P(n)\Big[(1-q)\,\mathcal{S}_n^{\mathrm{MR}}(c)+q\,\mathcal{S}_n^{\mathrm{FD}}(c)\Big],
  \label{eq:second_moment_decomp}
\end{equation}
and the diffusion coefficient is then approximated from Eq.~\eqref{eq:D_from_RL_exact} as
$D(c)\approx [\mathcal{S}(c)-\langle\Delta N\rangle^2]/N$.
The majority-rule contribution within this closure is
\begin{align}
  \mathcal{S}_n^{\mathrm{MR}}(c)
  & \approx \sum_{\ell=\lfloor n/2\rfloor+1}^{n} (n-\ell)^2 B_{n,\ell}(c)
   + \sum_{\ell=0}^{\lfloor (n-1)/2\rfloor} \ell^2 B_{n,\ell}(c),
  \label{eq:SnMR_explicit}
\end{align}
where the same binomial weight $B_{n,\ell}(c)$ defined above has been used. For the field-driven branch, within the same closure, if $\ell$ denotes the number of $+1$ agents before the update, we approximate $\Delta N=n-\ell$ when the group is collectively set to $+1$, and $\Delta N=-\ell$ when the group is collectively set to $-1$. Averaging over the binomial pre-update composition then gives
\begin{align}
  \mathcal{S}_n^{\mathrm{FD}}(c)
  &\approx p\sum_{\ell=0}^{n}(n-\ell)^2 B_{n,\ell}(c)
   +(1-p)\sum_{\ell=0}^{n}\ell^2 B_{n,\ell}(c) \nonumber \\
  &\approx nc\left(1-c\right)
   + n^2 \left[p\left(1-c\right)^2 + \left(1-p\right)c^2\right].
  \label{eq:SnFD_explicit}
\end{align}
Averaging Eqs.~\eqref{eq:SnMR_explicit}--\eqref{eq:SnFD_explicit} over $P(n)$ yields an approximate expression for $\mathcal{S}(c)$, and substituting it into Eq.~\eqref{eq:D_from_RL_exact} gives the corresponding approximation for the diffusion coefficient $D(c)$.

For $q>0$ and $0<p<1$, the consensus boundaries $c=0$ and $c=1$ are generally not absorbing, because the field-driven branch can reintroduce either opinion into a homogeneous configuration. Absorbing consensus persists only in the pure majority-rule limit $q=0$ or at the extreme biases $p=0$ and $p=1$. This distinction will be used below when interpreting the first-passage observables.

We emphasize that the diffusion approximation retains only the first two one-step moments within the annealed MF closure. It is therefore expected to be most accurate when typical jumps in $c$ remain small, while broad hyperedge-size distributions can generate finite-size corrections through rare large-group events. In the disordering-time problem, additional quantitative deviations may also arise from the drift-dominated time-of-flight approximation and the linearization near the stable mixed fixed point.

\subsection{\label{sec:sub_section_iii_ii}Criticality}
At the up-down symmetric point $p=1/2$, the balanced state $c=1/2$ is always a fixed point. The order--disorder transition occurs when this fixed point changes its linear stability.

Using the drift in Eq.~\eqref{eq:drift_general} and the antisymmetry $\Delta M_n(1-c)=-\Delta M_n(c)$, we expand $v(c)$ around the fixed point $c=1/2$. To leading order in the deviation $\delta c = c - 1/2$ (see Appendix~\ref{app:rates}), one obtains the linearized form
\begin{equation}
  v(c) \simeq
  \left[\left(1-q\right)\mathcal{A}_1(P)-q\,\langle n\rangle\right] \left(c-1/2\right),
  \label{eq:landau_expansion}
\end{equation}
where $\langle n\rangle = \sum_{n} nP(n)$ is the mean hyperedge size, and $\mathcal{A}_1(P)$ measures the average strength of the majority rule near the symmetric point. This coefficient is given by
\begin{equation}\label{eq:A_1}
    \mathcal{A}_1(P) = \sum_{n=3}^{n_{\max}} P(n)\,2^{2-n} S_n,
\end{equation}
where the combinatorial factor
$S_n=\sum_{\ell=\lfloor n/2\rfloor+1}^{n} (n-\ell)(2\ell-n)\binom{n}{\ell}$
represents the specific majority leverage for groups of size $n$. The subscript ``1" in $\mathcal{A}_1(P)$ indicates that this quantity is the coefficient of the linear term in the expansion of the drift around the symmetric fixed point $c=1/2$.
Physically, the first term in the bracket of Eq.~\eqref{eq:landau_expansion} describes the ordering tendency generated by majority-rule updates, while the term proportional to $q\,\langle n\rangle$ reflects the tendency of the field-driven branch to restore the unbiased value $c=1/2$ when $p=1/2$. Equivalently, $\mathcal{A}_1(P)$ quantifies how strongly majority updates amplify small imbalances away from $c=1/2$, whereas $q\,\langle n\rangle$ quantifies how strongly collective field-driven updates suppress such imbalances by reorienting the selected group without reference to its local composition.

The disordered fixed point $c=1/2$ loses stability when the slope $v'(1/2)$ becomes positive (i.e., when the drift pushes the system away from $1/2$). Setting the term in square brackets in Eq.~\eqref{eq:landau_expansion} to zero yields the critical field-driven update probability (probability threshold)
\begin{equation}
  q_c(P)
  = \frac{\mathcal{A}_1(P)}
         {\mathcal{A}_1(P)+\langle n\rangle}.
  \label{eq:qc_general}
\end{equation}
This expression highlights that $q_c(P)$ is determined by the competition between the average majority leverage $\mathcal{A}_1(P)$ and the relaxation induced by the collective field-driven branch. Physically, $q_c(P)$ marks the level of field-driven updating at which the linear stability of the symmetric mixed state changes: for $q<q_c(P)$ majority amplification dominates and the symmetric mixed state is unstable, whereas for $q>q_c(P)$ field-driven updates dominate and stabilize the mixed state around $c=1/2$. Hence, an increase in $q_c$ means that ordering is more robust against this noise mechanism, i.e., a stronger field-driven tendency is required to destroy macroscopic order.

It is important to stress that Eq.~\eqref{eq:qc_general} is specific to the up-down symmetric case $p=1/2$. For $p\neq 1/2$, the balanced state $c=1/2$ is no longer a fixed point, since the field-driven contribution in Eq.~\eqref{eq:drift_general} gives $v(1/2)=q\,\langle n\rangle (p-1/2)\neq 0$. In that case, the phase portrait is biased and the relevant stationary state, when it exists, is shifted to a location $c^\ast\neq 1/2$ determined by $v(c^\ast)=0$. Accordingly, the transition discussed in this subsection should be understood as the symmetric order--disorder transition at $p=1/2$. Away from symmetry, the same parameter $q$ still controls the relative importance of majority and field-driven updates, but the long-time dynamics is more appropriately described in terms of a biased ordered or mixed attractor rather than by the same critical point $q_c(P)$.

Specializing Eq.~\eqref{eq:qc_general} to the three hyperedge-size families yields explicit expressions for the critical field-driven update level in the symmetric case. For the $n$-uniform ensemble, we have
\begin{equation}
  \label{eq:crit_q_uniform}
  q_c(n)=\dfrac{2^{2-n}S_n}{2^{2-n}S_n+n}.
\end{equation}
Equation~\eqref{eq:crit_q_uniform} shows that $q_c(n)$ takes the same value for successive odd--even pairs, i.e., for $n$ and $n+1$ when $n$ is odd. This is because increasing the group size from odd $n$ to even $n+1$ introduces only tied configurations with equal numbers of $+1$ and $-1$ agents. Since ties produce no update in the present dynamics, they do not affect the linear stability of the symmetric state and hence leave $q_c$ unchanged. As a result, $q_c(n)$ increases in a stepwise manner, changing only when the group size moves to the next odd--even pair.

For the shifted geometric family, a compact expression is obtained in the thermodynamic limit. In this limit, Eq.~\eqref{eq:qc_general} becomes
\begin{equation}
  \label{eq:crit_q_geo}
  q_c(x)=\frac{(1-x)^2\sum_{n=3}^{\infty} x^{n-3}2^{2-n}S_n}
              {(1-x)^2\sum_{n=3}^{\infty} x^{n-3}2^{2-n}S_n + (3-2x)}.
\end{equation}
The corresponding mean group size is then $\langle n\rangle=(3-2x)/(1-x)$, so that the threshold $q_c$ is an increasing function of $\langle n\rangle$, indicating that larger interaction groups enhance majority amplification and therefore require a larger field-driven update probability to destroy order.

For the truncated power-law ensemble, the normalization constants cancel in Eq.~\eqref{eq:qc_general}, yielding
\begin{equation}
  \label{eq:crit_q_pl}
  q_c(\alpha)=
  \frac{\sum_{n=3}^{N} n^{-\alpha}2^{2-n}S_n}
       {\sum_{n=3}^{N} n^{-\alpha}2^{2-n}S_n + \sum_{n=3}^{N}n^{1-\alpha}}.
\end{equation}
In the thermodynamic limit $N\to\infty$, this expression converges to
\begin{equation}
  \label{eq:qc_pl_asymp}
  q_c(\alpha)\to
  \begin{cases}
    1, & \alpha\le 5/2,\\[6pt]
    \displaystyle
    \frac{\sum_{n=3}^{\infty} n^{-\alpha}2^{2-n}S_n}
         {\sum_{n=3}^{\infty} n^{-\alpha}2^{2-n}S_n+\zeta(\alpha-1,3)},
    & \alpha> 5/2,
  \end{cases}
\end{equation}
where $\zeta(a,b)$ denotes the Hurwitz zeta function. 
Thus, within the symmetric case $p=1/2$, a genuine order--disorder transition exists only for $\alpha>5/2$. In this regime, $q_c(\alpha)$ decreases monotonically with $\alpha$ and approaches the $n=3$ limit, $q_c=1/3$, as $\alpha\to\infty$, where the distribution becomes effectively $3$-uniform. By contrast, for $\alpha\le 5/2$, the system remains ordered for any $q<1$; equivalently, in the thermodynamic limit the symmetric mixed state becomes linearly stable only in the fully field-driven limit $q\to 1$.

\begin{figure}[t]
    \centering
    \includegraphics[width=\linewidth]{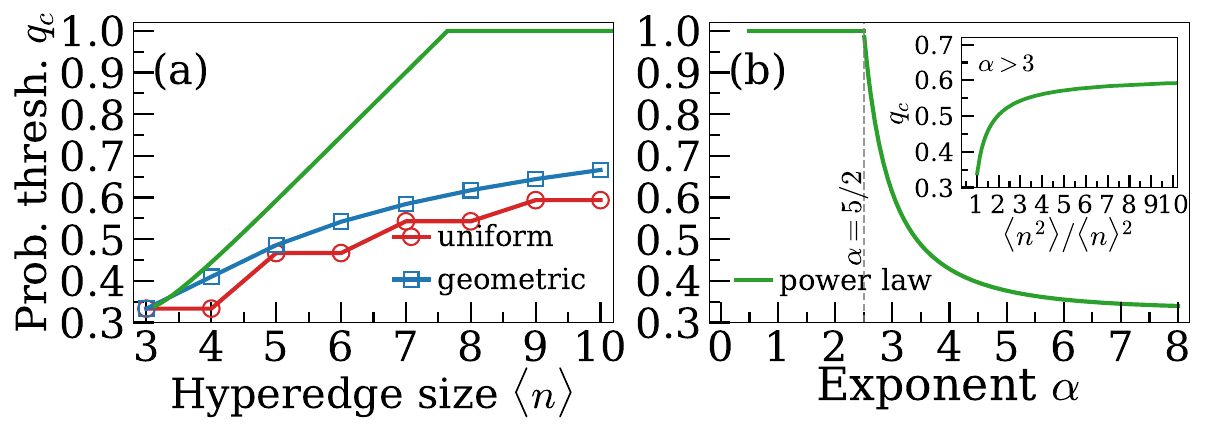}
    \caption{Phase diagram of the noisy majority-rule dynamics on annealed hypergraphs in the up--down symmetric case $p=1/2$. 
    (a) Probability threshold $q_c$ plotted versus the mean group size $\langle n\rangle$ for the $n$-uniform [Eq.~\eqref{eq:crit_q_uniform}], shifted geometric [Eq.~\eqref{eq:crit_q_geo}], and power-law [Eq.~\eqref{eq:qc_pl_asymp}] hyperedge-size distributions. For the $n$-uniform ensemble, $\langle n\rangle=n$. 
    (b) Probability threshold $q_c$ for the power-law ensemble as a function of the tail exponent $\alpha$. For $\alpha\le 5/2$, the thermodynamic-limit threshold reaches $q_c=1$, indicating suppression of the symmetric order--disorder transition except in the fully field-driven limit. For $\alpha>5/2$, a finite threshold $q_c<1$ is obtained; the vertical dashed line marks $\alpha=5/2$. The inset shows the finite-second-moment branch $\alpha>3$, replotted versus the reduced second moment $\langle n^2\rangle/\langle n\rangle^2$. The system is ordered for $q<q_c$ and disordered for $q>q_c$.}
    \label{fig:phase_diagram}
    \label{fig:critical_point}
\end{figure}

To help distinguish the role of the average group size from that of the distributional breadth, Fig.~\ref{fig:critical_point}(a) replots the symmetric threshold $q_c$ against the mean group size $\langle n\rangle$. For the $n$-uniform ensemble, this simply reduces to $\langle n\rangle=n$, while for the shifted geometric and power-law ensembles $\langle n\rangle$ is computed from the corresponding hyperedge-size distribution. This representation allows the three distribution families to be compared on the same horizontal scale and shows that the ordered phase generally becomes more robust as the average group size increases. At comparable $\langle n\rangle$, however, the power-law curve rises more steeply than the shifted-geometric one because its heavier tail assigns more weight to rare large hyperedges, thereby enhancing the majority leverage more strongly and yielding a larger $q_c$.

For the power-law family, however, varying $\alpha$ changes not only the characteristic group size but also the breadth of $P(n)$. For this reason, Fig.~\ref{fig:critical_point}(b) shows $q_c$ as a function of $\alpha$, while the inset replots only the finite-second-moment branch against the reduced second moment $\langle n^2\rangle/\langle n\rangle^2$, which provides a convenient measure of group-size fluctuations. The inset is restricted to $\alpha>3$, where the reduced second moment is finite; hence it does not include the regime $\alpha\le 5/2$, for which $q_c=1$ in the thermodynamic limit. This complementary representation shows that the variation of $q_c$ in the power-law ensemble reflects changes not only in the typical group size but also in the breadth of the group-size distribution.

Equations~\eqref{eq:crit_q_uniform}--\eqref{eq:qc_pl_asymp} show how higher-order interaction patterns control the tolerance of ordering to field-driven noise through the symmetric threshold $q_c$. In particular, distributions biased toward larger groups increase the majority leverage $\mathcal{A}_1(P)$, so a larger field-driven update probability is required to stabilize the mixed state, yielding a larger $q_c$ [Fig.~\ref{fig:critical_point}(a)]. For truncated power-law ensembles, small exponents $\alpha$ place substantial weight on large hyperedges, driving $q_c(\alpha)$ toward unity, whereas for large $\alpha$ the distribution is dominated by small groups and $q_c$ approaches the $n=3$ limit. In this sense, heavy-tailed $P(n)$ makes disordering harder: rare but large hyperedges produce strong majority amplification, so the system can remain ordered even when field-driven updates are frequent.

The case $p \neq 1/2$ further clarifies the interplay between majority amplification and the field-driven branch.
At the extreme biases $p = 1$ and $p = 0$, the field-driven branch acts constructively with majority rule toward the corresponding favored consensus state, namely $c = 1$ or $c = 0$. For intermediate biases $0<p<1$, the field-driven branch still acts as a symmetry-breaking bias and shifts the stable attractor toward the favored opinion; in this sense, it enhances asymmetric ordering.
However, unlike the extreme-bias cases $p=0$ and $p=1$, it does not make the corresponding consensus boundary absorbing. In this intermediate-bias regime, the interaction between the two mechanisms is state dependent: the field-driven term drives the dynamics toward $c=p$, whereas the majority contribution pushes the system away from $c=1/2$. Hence the two contributions reinforce each other only over part of the interval $0<c<1$ and compete elsewhere. The corresponding long-time dynamics should therefore be interpreted in terms of relaxation toward a bias-selected ordered or mixed attractor, depending on the existence and stability of the biased fixed point, rather than in terms of the symmetric threshold $q_c$ itself.

\section{\label{sec:section_iv} Consensus Time}
\subsection{\label{sec:sub_section_iv_i}Pure majority-rule regime}

The consensus time or ordering time is the mean first-passage time (MFPT) for the coarse-grained fraction $c(\tau)$ of $+1$ opinions to reach an absorbing consensus configuration, i.e., $c=1$ (all agents in state $+1$) or $c=0$ (all agents in state $-1$), starting from a partially ordered initial condition. In the pure majority-rule dynamics ($q=0$), updates are governed solely by the local majority within the sampled hyperedge, so the characteristic time scale is controlled by the hyperedge size or its distribution $P(n)$. In the thermodynamic limit, the fixed point at $c=1/2$ is linearly unstable; trajectories with $c_0>1/2$ flow to $c=1$, whereas those with $c_0<1/2$ flow to $c=0$. Dynamically, consensus is drift-dominated away from the separatrix at $c=1/2$, while demographic fluctuations set the width of the initial neighborhood from which the system escapes, leading to the characteristic logarithmic growth of the MFPT with $N$ in well-mixed settings.

A standard analysis employs the backward Kolmogorov description for the one-dimensional process $c(\tau)$ with drift $v(c)$ and diffusion $D(c)$. The MFPT from $c_0$ to an absorbing boundary $c_b$ (assuming a reflecting boundary at $c_a$; e.g., $c_a=1/2$, $c_b=1-1/N$) solves $v(c)T'(c)+\frac{1}{2}D(c)T''(c)=-1$ with $T'(c_a)=0$ and $T(c_b)=0$. The solution is given by the quadrature~\cite{gardiner2009stochastic,van1992stochastic}
\begin{equation}
  T(c_0)= \int_{c_0}^{c_b}dy\,\exp(-\Psi(y))\int_{c_a}^{y}dx\,\dfrac{2\exp(\Psi(x))}{D(x)},
  \label{eq:mfpt_quadrature}
\end{equation}
where $\Psi(c)=\int_{c_a}^{c}\left[2v(z)/D(z)\right]\,dz$. All dependence on the higher-order structure enters through $v(c)$ and $D(c)$, which are determined by the majority leverage of size-$n$ groups averaged over $P(n)$.

The quadrature in Eq.~\eqref{eq:mfpt_quadrature} does not admit a closed form for general $P(n)$. However, in the large-$N$ regime, when the dynamics is predominantly drift driven, the MFPT is well approximated by its leading deterministic contribution. In this case, the main contribution comes from the time of flight of the coarse-grained dynamics $dc/d\tau = v(c)$, yielding
\begin{equation}
  T \approx \int_{c_i}^{c_f} \frac{dc}{v(c)}.
  \label{eq:tof_leading}
\end{equation}
Here, $c_i$ and $c_f$ are chosen along a deterministic trajectory such that $v(c)$ has the same sign as $(c_f-c_i)$, ensuring $T>0$. Physically, Eq.~\eqref{eq:tof_leading} indicates that in a well-mixed population the ordering time is controlled primarily by the majority-induced drift, while fluctuations mainly set the width of the boundary layer from which the system escapes.

For initially balanced configurations, we regularize the unstable fixed point at $c=1/2$ by displacing the initial condition by its natural finite-size fluctuation, $c_i=1/2+\delta$ with $\delta \sim N^{-1/2}$, and we set the absorbing threshold at $c_f=1-1/N$. The shift $\delta$ reflects the typical demographic imbalance arising from random initial conditions, representing the minimal bias that majority pressure can amplify~\cite{van1992stochastic}. The choice $c_f=1-1/N$ avoids the singular boundary at $c=1$ while capturing the ``last minority'' conversion. This prescription prevents long, uninformative sojourns near the unstable point and isolates the genuine $N$-dependence of the consensus time.

Evaluating the integral in Eq.~\eqref{eq:tof_leading} (see Appendix~\ref{appendixB} for details) yields the robust scaling
\begin{equation}
  T_{\mathrm{cons}} \sim \mathcal{B}\,\ln N,
  \label{eq:T_logN_generic}
\end{equation}
with a prefactor $\mathcal{B}$ that has a clear physical interpretation. The logarithm arises from two stages: (i) the escape from the vicinity of the unstable state ($c \simeq 1/2$), and (ii) the approach to full consensus where the remaining minority is swept out. The first stage is determined by the local slope of the majority drift near $c=1/2$, while the second is controlled by the typical group size near the absorbing boundary.

Specializing to the annealed $n$-uniform ensemble, the two contributions combine into the prefactor
\begin{equation}
  \label{eq:prefac_uniform_main}
  \mathcal{B}(n)=\frac{1}{n}+\frac{2^{n-3}}{S_n},
\end{equation}
where $S_n$ quantifies the net restoring effect of strict majorities in size-$n$ groups. Equation~\eqref{eq:prefac_uniform_main} holds for both odd and even $n$: for even $n$, tie events correspond to $\ell=n/2$ and are excluded by construction in $S_n$, yielding no flips. The term $1/n$ reflects the final elimination of the last minority, whereas the term $2^{n-3}/S_n$ accounts for the slow passage through the nearly balanced region.

For the shifted geometric family, the prefactor also takes a compact form in the thermodynamic limit. In this limit, one obtains
\begin{equation}
  \label{eq:prefac_geo_main}
  \mathcal{B}(x)
  =\frac{1-x}{3-2x}
  +\frac{1}{\sum_{n\ge 3}(1-x)x^{n-3}2^{3-n} S_n}.
\end{equation}
The first term is $1/\langle n\rangle$ with the thermodynamic-limit mean $\langle n\rangle=(3-2x)/(1-x)$, and therefore represents the final sweep toward consensus. The second term measures the inverse effective majority leverage near $c\simeq 1/2$.

\begin{figure*}[tb!]
    \centering
    \includegraphics[width=\linewidth]{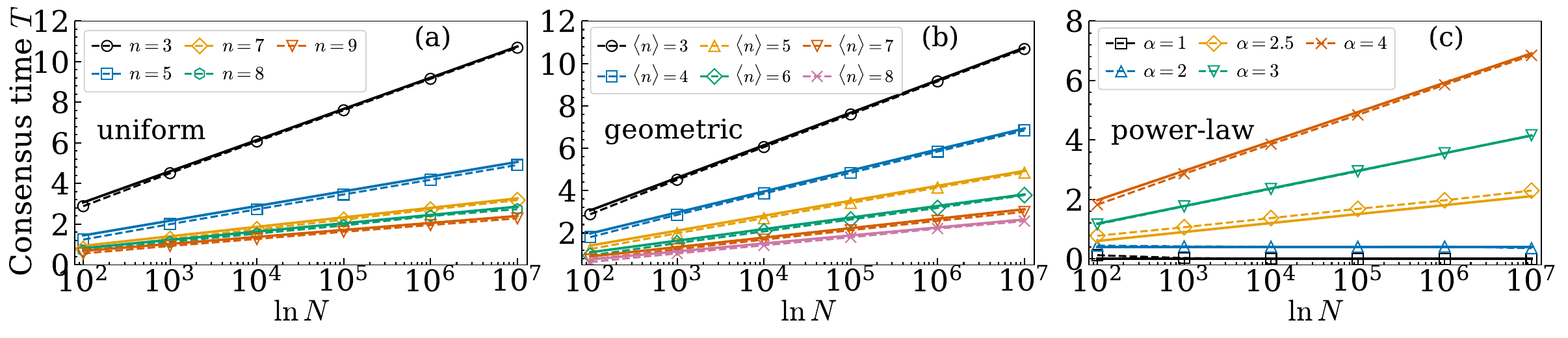}
    \caption{Consensus time $T$ versus system size $N$ in the pure majority-rule regime ($q=0$), starting from the balanced initial condition $c_0=1/2$, for three hyperedge-size distributions: (a) annealed $n$-uniform for several $n$; (b) shifted geometric for several mean sizes $\langle n\rangle$; and (c) truncated power law for several tail exponents $\alpha$. Solid curves show the MF prediction from Eq.~\eqref{eq:T_logN_generic} with the corresponding prefactors. Symbols connected by dashed lines are MC data averages over $10^{5}$ runs. The data confirm the predicted crossover: logarithmic growth with a distribution-dependent prefactor for the $n$-uniform and shifted-geometric ensembles, as well as for truncated power-law ensembles with $\alpha>2$, and saturation for sufficiently heavy-tailed power laws with $\alpha\le 2$.}
    \label{fig:cons_time_noiseless}
\end{figure*}

For a truncated power law with cutoff $n_{\max}=N$, the finite-$N$ prefactor reads
\begin{equation}
  \label{eq:B_pl_finiteN}
  \mathcal{B}(N,\alpha)
  =\frac{\sum_{n=3}^{N} n^{-\alpha}}{\sum_{n=3}^{N} n^{1-\alpha}}
  +\frac{\sum_{n=3}^{N} n^{-\alpha}}{\sum_{n=3}^{N} n^{-\alpha}2^{3-n}S_n}.
\end{equation}
In the thermodynamic limit, the asymptotic behavior depends on $\alpha$:
\begin{equation}
  \label{eq:B_pl_limits}
  \mathcal{B}(\alpha)=
  \begin{cases}
    0, & \alpha \le 2,\\[6pt]
    \dfrac{\zeta(\alpha,3)}{\zeta(\alpha-1,3)}, & 2<\alpha\le 5/2,\\[10pt]
    \dfrac{\zeta(\alpha,3)}{\zeta(\alpha-1,3)}
    +\dfrac{\zeta(\alpha,3)}{\sum_{n=3}^{\infty} n^{-\alpha}2^{3-n}S_n}, & \alpha> 5/2.
  \end{cases}
\end{equation}
For $\alpha\le 2$, the prefactor vanishes as $N\to\infty$, signaling that the consensus time grows slower than $\ln N$ (or saturates). For $\alpha>2$, the prefactor is strictly positive, and $T_{\mathrm{cons}}$ recovers the logarithmic form $T_{\mathrm{cons}}\sim \mathcal{B}(\alpha)\,\ln N$. Note that for $2 < \alpha \le 5/2$, only the first term in Eq.~\eqref{eq:B_pl_finiteN} survives in the limit, while for $\alpha > 5/2$, both contributions remain finite. These predictions are summarized in Fig.~\ref{fig:cons_time_noiseless}, which shows the system-size dependence of $T_{\mathrm{cons}}$ for the three families of hyperedge-size distributions considered here. In particular, the data confirm logarithmic growth with a distribution-dependent prefactor for the $n$-uniform and shifted-geometric ensembles, as well as for truncated power-law ensembles with $\alpha>2$, whereas sufficiently heavy-tailed power laws ($\alpha\le 2$) exhibit saturation or sublogarithmic growth.

The trends visible in Fig.~\ref{fig:cons_time_noiseless} are consistent with the analytical prefactors in Eqs.~\eqref{eq:prefac_uniform_main}--\eqref{eq:B_pl_limits}. Shifting $P(n)$ toward larger group sizes (which typically increases the truncated mean $\langle n\rangle_N$ and enhances the weight of large-$n$ events) strengthens the effective majority leverage and thus speeds up ordering. For the $n$-uniform case, $B(n)$ decreases monotonically with $n$, yielding the longest consensus time at $n = 3$ [$T_{\max}(N) = (2/3)\ln N$] and tending to zero as $n \to \infty$. This monotonicity holds for both odd and even $n$, since tie events for even $n$ contribute no flips and are excluded by construction in the strict-majority sums defining $S_n$. Similarly, for the geometric distribution, $B(x)$ decreases with the mean size $\langle n\rangle$. This suggests that larger groups exhibit stronger majority-restoring tendencies, correcting deviations from consensus more efficiently.

The truncated power-law ensemble exhibits the
strongest sensitivity to group sizes.
Here, varying the exponent $\alpha$ changes both the tail weight and the truncated mean size $\langle n\rangle_N(\alpha)$; therefore the $\alpha$-dependence of $T_{\rm cons}$ reflects the combined effect of suppressing/allowing large hyperedges and changing $\langle n\rangle_N$.
Heavy tails ($\alpha \le 2$)
assign significant weight to very large hyperedges, which
can align macroscopic fractions of the system in a single update,
causing $T_{\rm cons}$ to saturate or grow sublogarithmically.
Conversely, for lighter tails ($\alpha > 2$), the
relevant moments are finite and the logarithmic scaling holds.
As $\alpha \to \infty$, the size distribution becomes effectively
3-uniform, and the prefactor approaches the $n = 3$ limit, $B \to 2/3$.

\begin{figure}[tb!]
    \centering
    \includegraphics[width=\linewidth]{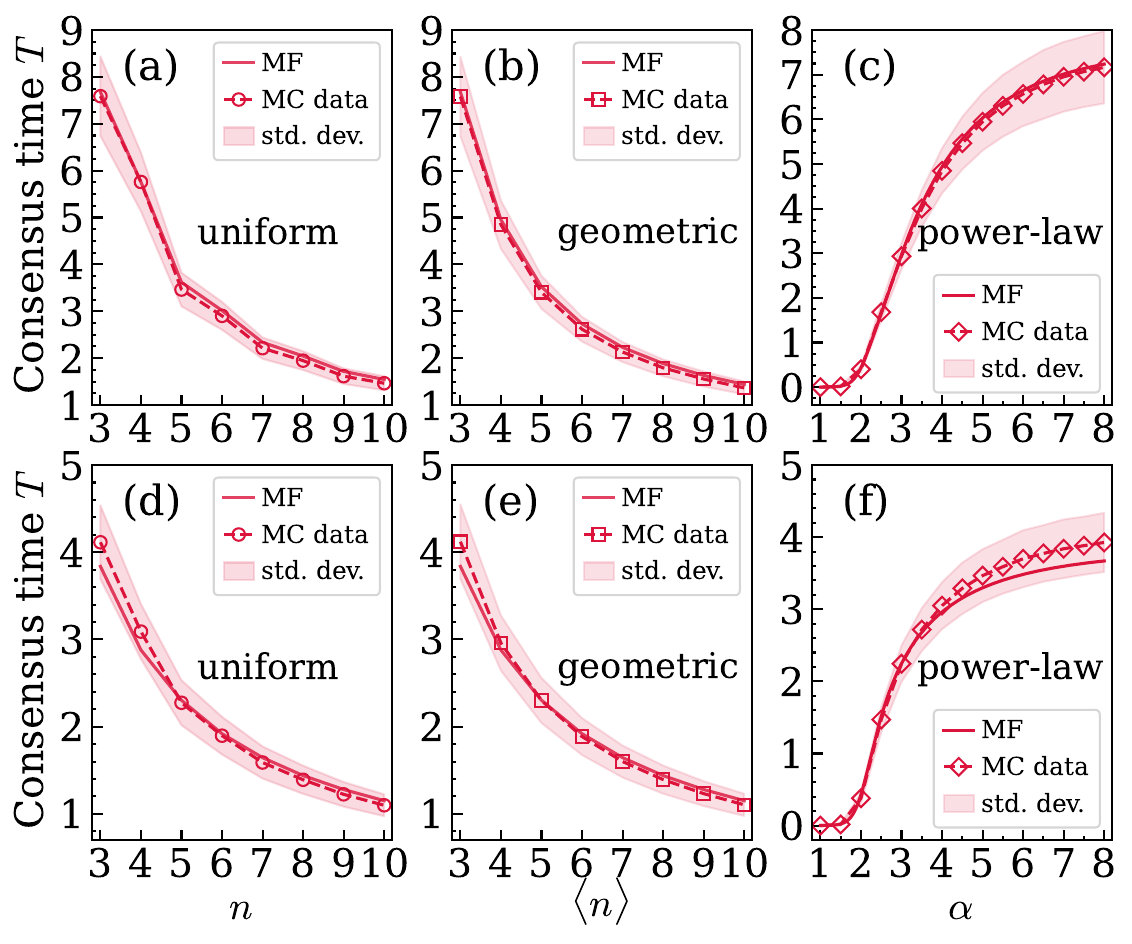}
    \caption{Consensus time $T$ at fixed system size $N=10^{5}$ in the pure majority-rule regime ($q=0$) for three hyperedge-size distributions under two initial conditions. Panels (a)--(c): balanced start $c_0=0.5$. Panels (d)--(f): unbalanced start $c_0=0.7$. (a),(d) Annealed $n$-uniform (varying $n$); (b),(e) shifted geometric (varying $\langle n\rangle$); and (c),(f) truncated power law (varying $\alpha$). Solid curves show the MF predictions, using Eqs.~\eqref{eq:prefac_uniform_main}--\eqref{eq:B_pl_limits} for panels (a)--(c) and Eqs.~\eqref{eq:unif_cons_noiseless}--\eqref{eq:pl_cons_noiseless} for panels (d)--(f). Symbols connected by dashed lines are MC averages over $10^{5}$ realizations. Shaded bands indicate one standard deviation.}
    \label{fig:cons_time_noiseless_g}
\end{figure}

For fixed system size $N$, these dependencies are visualized in Fig.~\ref{fig:cons_time_noiseless_g}.
$T_{\rm cons}$ decreases monotonically with increasing group size (uniform/geometric)
and increases with the exponent $\alpha$ (power law),
consistent with the shift of probability weight toward smaller hyperedges and the concomitant decrease of $\langle n\rangle_N(\alpha)$ for our truncation.
The MC data [Fig.~\ref{fig:cons_time_noiseless_g}(a)–(c)] closely match these analytical predictions. The power-law result may also be reparametrized in terms of the moments of $P(n)$, as shown in Fig.~\ref{fig:app_powerlaw_moments} of Appendix~\ref{sub:subsub_pl}. This representation leads to the same qualitative conclusion: larger effective interaction sizes and broader tails are associated with faster consensus, although the detailed behavior is still governed by the full form of $P(n)$ rather than by these moments alone.

When the initial condition is displaced by an $O(1)$ amount from the unstable fixed point, the dynamical bottleneck at $c=1/2$ is bypassed. Consequently, the mean consensus time is determined solely by the final approach to the absorbing boundary. To leading order, one recovers the logarithmic scaling of Eq.~\eqref{eq:T_logN_generic}, but with a prefactor that lacks the contribution from the unstable fixed point. This modified prefactor depends only on the statistics of the hyperedge sizes:

For the annealed $n$-uniform distribution, the prefactor simplifies to
\begin{equation}
  \label{eq:unif_cons_noiseless}
  \mathcal{B}(n)=\frac{1}{n},
\end{equation}
indicating that consensus accelerates inversely with group size. For the shifted geometric distribution,
\begin{equation}
  \label{eq:geo_cons_noiseless}
  \mathcal{B}(x)=\frac{1-x}{3-2x},
\end{equation}
which is a decreasing function of the mean size $\langle n\rangle=(3-2x)/(1-x)$.

For the truncated power law, with natural cutoff $n_{\max}=N$, the prefactor exhibits three distinct regimes:
\begin{equation}
  \label{eq:pl_cons_noiseless}
  \mathcal{B}(N,\alpha)\simeq
  \begin{cases}
    0, & \alpha<2,\\[4pt]
    \dfrac{\zeta(2,3)}{\ln N}, & \alpha=2,\\[10pt]
    \dfrac{\zeta(\alpha,3)}{\zeta(\alpha-1,3)}, & \alpha>2.
  \end{cases}
\end{equation}
Thus, genuine logarithmic scaling persists only for $\alpha>2$. At the marginal point $\alpha=2$, the logarithmic growth of $T_{\mathrm{cons}}$ is canceled by the vanishing prefactor ($\mathcal{B} \sim 1/\ln N$), resulting in an $N$-independent saturation. For $\alpha<2$, the consensus time is similarly $O(1)$.

These MF predictions are confirmed by the simulations in Fig.~\ref{fig:cons_time_noiseless_g}(d)--(f): $T_{\mathrm{cons}}$ decreases with increasing $n$ or $\langle n\rangle$, and increases with $\alpha$. This is consistent with the physical intuition that heavy-tailed distributions (small $\alpha$) facilitate rapid consensus by frequently sampling large, highly efficient groups.

\subsection{\label{sec:sub_section_iv_ii} Field-driven regime}

In the field-driven regime ($q>0$), the long-time behavior depends on both the update probability $q$ and the bias strength $p$. At the symmetric point $p=1/2$, sufficiently large $q$ stabilizes the mixed state at $c=1/2$, and the corresponding first-passage observable is the disordering time discussed in Sec.~\ref{sec:section_v}. For $p\neq 1/2$, the up--down symmetry is explicitly broken and the stable attractor, when it exists, is shifted to a biased location $c^\ast$ determined by $v(c^\ast)=0$.

For intermediate biases $0<p<1$, the states $c=0$ and $c=1$ are not absorbing whenever $q>0$, because the field-driven branch can reintroduce either opinion into a homogeneous configuration. Therefore, a strict consensus time to $c=0$ or $c=1$ is not well defined in this regime. The appropriate quantity would instead be a relaxation time toward the corresponding biased attractor. However, a detailed analysis of this biased-relaxation problem is beyond the scope of the present work, which focuses on absorbing consensus in the pure majority-rule limit and at extreme biases, and on the symmetric disordering time at $p=1/2$.

By contrast, in the extreme-bias limits $p=0$ and $p=1$, one of the two boundary consensus states remains absorbing. In these cases, the field-driven branch and the majority rule act constructively toward the selected consensus state, allowing the consensus time to be analyzed. Socially, this limit can be interpreted as a fully aligned external information environment, such as a strongly one-sided media campaign or institutional signal, that consistently favors one opinion and prevents the reintroduction of the opposite state~\cite{carro2016noisy,muslim2024mass}.

For $p\in\{0,1\}$, the dynamical bottleneck at $c=1/2$ vanishes. The consensus time is therefore controlled solely by the relaxation of the boundary layer near the absorbing state. Consequently, the leading asymptotic behavior coincides with that of the pure majority-rule regime starting from unbalanced conditions, recovering the scaling $T_{\mathrm{cons}}\sim \mathcal{B}\,\ln N$ with the prefactors given in Eqs.~\eqref{eq:unif_cons_noiseless}--\eqref{eq:pl_cons_noiseless} (see Appendix~\ref{appendixC} for details). The system bypasses the nearly balanced bottleneck and proceeds directly via the progressive erosion of the minority.

Remarkably, in this limit, the logarithmic prefactor $\mathcal{B}$ becomes independent of the field-driven update probability $q$. This universality arises because, at extreme bias, the local majority rule and the directional bias $p$ act constructively. Near the absorbing boundary (e.g., $c\to 1$ with $p=1$), a selected hyperedge containing minority agents will convert them to the majority state regardless of whether the update is driven by peer pressure (majority rule) or by the external field. In other words, for $p=1$ (respectively $p=0$), both update branches deterministically favor the same boundary state $c=1$ (respectively $c=0$). This fully constructive interplay is special to the extreme-bias case and should not be extrapolated to intermediate $0<p<1$, where the field-driven term pulls the dynamics toward $c=p$ and may therefore reinforce or oppose the majority contribution depending on the current state $c$. Mathematically, near the selected absorbing boundary, the combined drift scales as $v(c)\simeq \langle n\rangle(1-c)$ for $p=1$ and $v(c)\simeq -\langle n\rangle c$ for $p=0$, equivalently $v(c)\simeq \langle n\rangle(p-c)$ in the corresponding boundary layer, leading to identical relaxation times. This robustness of the consensus time against the field-driven update probability $q$ mirrors analogous findings in our previous work~\cite{muslim2025ordering}.

\section{\label{sec:section_v} Disordering Time}

In this section, we restrict attention to the up--down symmetric case $p=1/2$, for which the mixed state is centered at $c=1/2$ and the symmetric threshold $q_c$ derived in Sec.~\ref{sec:sub_section_iii_ii} is well defined. For $p\neq 1/2$, the stable attractor, when it exists, is shifted away from $c=1/2$, and the relevant long-time dynamics is more appropriately described in terms of relaxation toward a biased attractor rather than by the disordering time defined below.

We define the disordering time $T_{\mathrm{dis}}$ as the MFPT required for the system to reach the neighborhood of the symmetric mixed state ($c=1/2$) starting from a fully ordered configuration. This observable is physically meaningful in the up-down symmetric case $p=1/2$ within the disordered regime $q>q_c$, where the fixed point at $c=1/2$ is linearly stable. In this regime, $T_{\mathrm{dis}}$ quantifies the robustness of consensus against noise: it measures how long an initially ordered population resists erosion under the combined action of local majority updates and field-driven fluctuations. Because the interacting group is resampled uniformly at random at each update, the dynamics remains annealed MF, which explains why the analytical and MC results stay close overall.

As in the consensus-time analysis, we adopt the backward FP picture and apply the drift-dominated approximation. The MFPT corresponds to the deterministic time of flight $dc/d\tau=v(c)$ integrated from the ordered boundary layer ($c \approx 1$) into the basin of attraction of the stable fixed point $c = 1/2$. Based on the linearization in Eq.~\eqref{eq:landau_expansion}, the drift close to $c=1/2$ can be written as
$v(c)\simeq -\lambda\left(c-1/2\right)$ with the relaxation rate $\lambda \equiv q\,\langle n\rangle - (1-q)\mathcal{A}_1$,
which is strictly positive in the disordered phase $q>q_c$. The integration is cut off at $c_f=1/2+O(N^{-1/2})$ to reflect the width of stationary fluctuations. This logarithmic approach to the fixed point dominates the integral, yielding the scaling
\begin{equation}
  \label{eq:polar_analytic}
  T_{\mathrm{dis}}\sim \mathcal{B}\,\ln N,
\end{equation}
with a prefactor $\mathcal{B}=1/(2\lambda)$.
Note that $\mathcal{B}$ diverges as $q \to q_c^+$, signaling the critical slowing down of the relaxation process. Compared with the consensus-time problem, however, the disordering problem is intrinsically more sensitive to finite-$N$ fluctuations because the target region is the fluctuation-dominated neighborhood of a stable interior fixed point rather than an absorbing boundary. This makes the asymptotic MFPT estimate quantitatively more delicate, even when the leading logarithmic scaling remains correct.

For the annealed $n$-uniform ensemble, substituting the explicit rate $\lambda$ into the prefactor yields (see Appendix~\ref{app:disordering_time} for details)

\begin{equation}\label{eq:polar_time_unif}
\mathcal{B}(n,q)=\Big[2\,q\,n-\left(1-q\right)2^{3-n}S_n\Big]^{-1}.
\end{equation}

Similarly, for the shifted geometric family, a compact expression is obtained in the thermodynamic limit. Summing over the group sizes then yields
\begin{equation}\label{eq:polar_time_geo}
\mathcal{B}(x,q)=\left[2\,q\,\frac{3-2x}{1-x}-\left(1-q\right)\left(1-x\right)\Phi(x)\right]^{-1},
\end{equation}
where $\Phi(x)=\sum_{n=3}^{\infty}x^{n-3}2^{3-n}S_n$ converges absolutely for $|x|<1$.

For the truncated power-law ensemble, the relaxation rate becomes $N$-dependent because the effective majority leverage diverges for $\alpha\le 5/2$. To make this explicit, we write
\begin{equation}
  \lambda_N \equiv q\,\langle n\rangle_N - (1-q)\mathcal{A}_{1,N}(\alpha),
  \label{eq:lambdaN_pl}
\end{equation}
where $  \mathcal{A}_{1,N}(\alpha)\equiv \sum_{n=3}^{N}P_N(n)2^{2-n}S_n$ and $\langle n\rangle_N$ are evaluated with the finite-size cutoff $n_{\max}=N$. The drift-dominated estimate then reads
\begin{equation}
  T_{\mathrm{dis}}(N;\alpha,q)\sim \frac{\ln N}{2\lambda_N},
  \label{eq:Tdis_pl_general}
\end{equation}
provided $\lambda_N>0$, i.e., provided that $c=1/2$ is linearly stable at the chosen $(N,\alpha,q)$.

The interpretation is as follows. For $\alpha>5/2$, both $\langle n\rangle_N$ and $\mathcal{A}_{1,N}(\alpha)$ converge, so $\lambda_N\to \lambda_\infty$ and Eq.~\eqref{eq:Tdis_pl_general} yields the genuine logarithmic growth $T_{\mathrm{dis}}\sim [2\lambda_\infty]^{-1}\ln N$. For $\alpha\le 5/2$, $\mathcal{A}_{1,N}(\alpha)$ diverges with $N$, implying $q_c(N)\to 1$ and hence the disordered regime requires $q$ to approach $1$ as $N$ increases; at fixed absolute $q<1$, the fixed point $c=1/2$ remains unstable asymptotically, and the disordering MFPT (as defined here) is not the appropriate descriptor of the long-$N$ dynamics. In finite systems, one may still observe rapid relaxation toward coexistence for $q$ sufficiently close to $1$, in which case Eq.~\eqref{eq:Tdis_pl_general} remains the correct starting point, with $\lambda_N$ set by the finite-size cutoff.

A useful consistency check is the pure field-driven limit $q=1$, where majority updates are absent. In this case, $\lambda=\langle n\rangle$ and the prefactor reduces to $\mathcal{B}=1/(2\langle n\rangle)$. This implies that disordering proceeds twice as fast (i.e., with half the time prefactor) as consensus formation from unbalanced initial conditions in the pure majority-rule regime [cf. Eqs.~\eqref{eq:unif_cons_noiseless}--\eqref{eq:pl_cons_noiseless}]. For instance, for the shifted geometric distribution, one recovers $\mathcal{B}(1,x)=(1-x)/[2(3-2x)]$. The factor of two originates from the different finite-size cutoffs: approaching an absorbing boundary requires reducing the minority to $O(1)$ agents (a distance $\sim 1/N$), whereas approaching the stable coexistence fixed point only requires reaching the scale of stationary fluctuations around $c=1/2$ (a distance $\sim N^{-1/2}$).

\begin{figure*}[tb!]
    \centering
    \includegraphics[width=\linewidth]{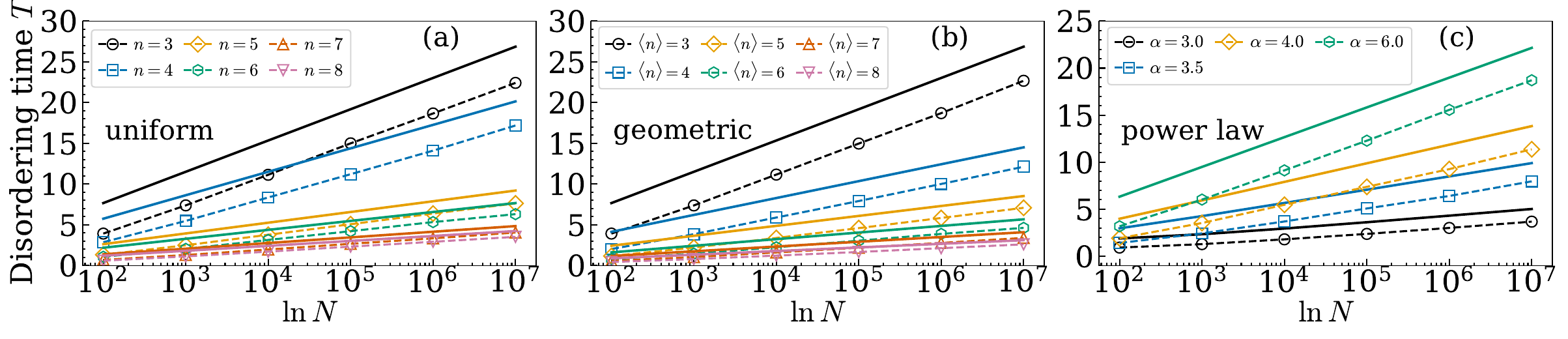}
    \caption{Disordering time $T_{\mathrm{dis}}$ versus system size $N$, starting from the fully ordered state $c_0=1$, for three hyperedge-size distributions in the up--down symmetric case $p=1/2$: (a) annealed $n$-uniform for various $n$; (b) shifted geometric for various $\langle n\rangle$; and (c) truncated power law for various $\alpha$. The data are shown at a fixed relative distance from the symmetric threshold, $q=\kappa q_c$ with $\kappa=1.2$. Solid curves show the MF prediction from Eq.~\eqref{eq:Tdis_kappa}, and symbols connected by dashed lines denote MC averages over $10^{5}$ runs. The results confirm the predicted logarithmic growth in the disordered phase ($q>q_c$), including power-law ensembles with $\alpha>5/2$.}
    \label{fig:dis_time}
\end{figure*}

To compare different hyperedge-size distributions on an equal footing, it is convenient to parameterize the distance from criticality as $q=\kappa q_c$ with $\kappa>1$.
Here and throughout this section, $q_c$ refers exclusively to the symmetric threshold at $p=1/2$.
This protocol is meaningful whenever $q_c(P)<1$ (in particular for the uniform and shifted-geometric ensembles, and for the truncated power-law ensemble with $\alpha>5/2$ in the thermodynamic limit).
This guarantees that the system is in the disordered phase regardless of the specific $P(n)$. Substituting $q=\kappa q_c$ into the linear relaxation rate $\lambda$ and using the definition of $q_c$ yields $\lambda = (\kappa-1)\mathcal{A}_1(P)$. Consequently, the disordering time obeys the compact scaling form
\begin{equation}
  T_{\mathrm{dis}}(N,\kappa)\sim\frac{\ln N}{2(\kappa-1)\mathcal{A}_1(P)}.
  \label{eq:Tdis_kappa}
\end{equation}
The parameter $\kappa$ controls the proximity to the threshold: as $\kappa \to 1^+$, the system exhibits critical slowing down with divergence $(\kappa-1)^{-1}$. The factor $\mathcal{A}_1(P)$ quantifies the structural stability provided by the hypergraphs. At a fixed relative distance $\kappa$, a larger $\mathcal{A}_1(P)$ implies a larger critical threshold $q_c$, and hence a larger absolute field-driven update probability $q=\kappa q_c$, leading to faster relaxation toward coexistence. Thus, the scaling form in Eq.~\eqref{eq:Tdis_kappa} is expected to capture the leading behavior well. Its quantitative accuracy, however, can be slightly lower than in the consensus-time problem because the disordering-time estimate relies on additional asymptotic steps, especially the drift-dominated time-of-flight approximation and the linearization near the stable mixed fixed point. For broad hyperedge-size distributions, rare large-group events can further enhance finite-size corrections by producing comparatively large jumps in $c$. The scaling form in Eq.~\eqref{eq:Tdis_kappa} is well supported by MC data for all three families of $P(n)$ as shown in  Fig.~\ref{fig:dis_time}.

For the $n$-uniform and shifted-geometric ensembles, $T_{\mathrm{dis}}$ decreases as the characteristic group size increases. This trend follows from the increase of $q_c$ with group size; consequently, at fixed $\kappa$, the corresponding absolute field-driven update probability $q=\kappa q_c$ is larger, leading to faster relaxation toward the disordered state.

\begin{figure}[tb!]
    \centering
    \includegraphics[width=\linewidth]{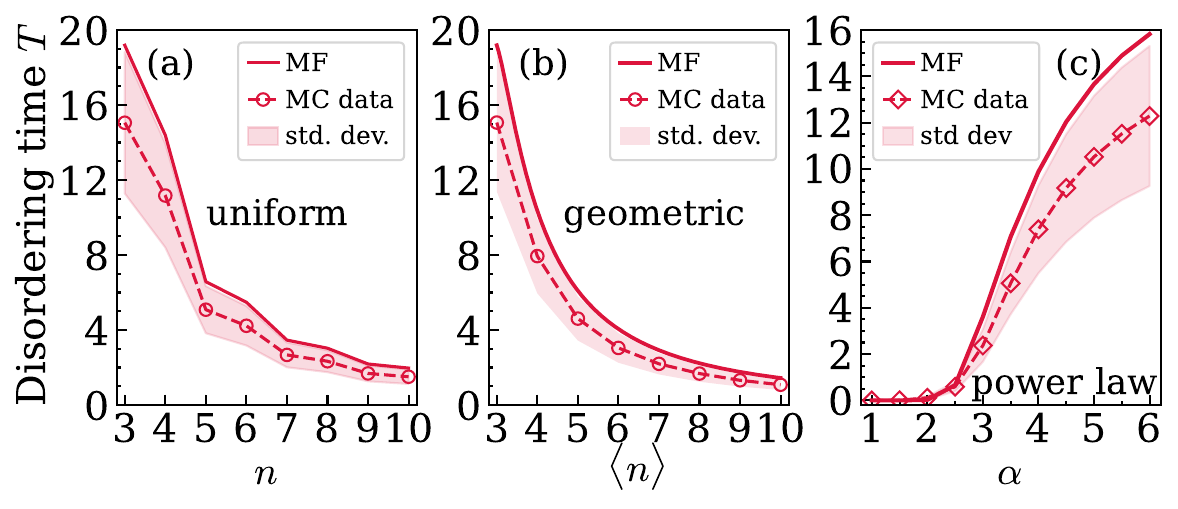}
    \caption{
    Disordering time $T_{\mathrm{dis}}$ at a fixed relative distance from the symmetric threshold, $\kappa=1.2$, in the up--down symmetric case $p=1/2$. Results are shown for systems of size $N=10^5$, with MC data averaged over $10^5$ independent realizations. Panels correspond to (a) the annealed $n$-uniform ensemble, (b) the shifted geometric ensemble, and (c) the truncated power-law ensemble. Solid lines show the mean-field prediction from Eq.~\eqref{eq:Tdis_kappa}, dashed lines with symbols denote MC results, and shaded regions indicate one standard deviation. For the truncated power-law ensemble, varying $\alpha$ changes both the tail weight and the truncated mean size $\langle n\rangle_N(\alpha)$, so the observed $\alpha$ dependence reflects the combined effect of suppressing large-$n$ events and reducing $\langle n\rangle_N$. The data confirm that, at fixed $\kappa$, systems with smaller effective group sizes, corresponding to small $n$, small $\langle n\rangle$, or larger $\alpha$, relax more slowly toward the disordered state.
    }
    \label{fig:polar_time}
\end{figure}

For power-law ensembles, the same intuition must be qualified because the effective moments of $P(n)$ may depend on $N$ under the cutoff, and the outcome becomes protocol dependent. At fixed absolute $q$, heavy tails lead to vanishing disordering times as $N\to\infty$ because the mean size $\langle n\rangle$ diverges and the field-driven component dominates the dynamics. At fixed relative distance $\kappa$ above threshold, $T_{\mathrm{dis}}$ increases with $\alpha$ (see Fig.~\ref{fig:polar_time}). In this constant-$\kappa$ comparison, lighter tails (large $\alpha$) imply smaller average group sizes and weaker majority leverage $\mathcal{A}_1(P)$, thereby slowing down the relaxation toward disorder. For the same reason, the diffusion approximation tends to perform best for the $n$-uniform and shifted-geometric ensembles, where typical jump sizes remain relatively controlled, and somewhat less accurately for broad power-law ensembles, where occasional large sampled groups produce more strongly non-diffusive increments.

\section{Exit probability}
\label{sec:exit_prob}

A central observable in opinion dynamics is the exit probability $E(c_0)$, defined as the probability that the dynamics reaches the all-$+1$ absorbing state ($c=1$) before the all-$-1$ state ($c=0$), starting from an initial fraction $c_0$ of $+1$ agents. In the diffusion approximation for the coarse-grained density $c(\tau)$, $E(c)$ satisfies the backward FP boundary-value problem~\cite{gardiner2009stochastic, risken1996fokker}
\begin{equation}
  v(c)\,E'(c)+\tfrac{1}{2}D(c)\,E''(c)=0,
  \label{eq:FP_exit_main}
\end{equation}
with absorbing boundary conditions $E(0)=0$ and $E(1)=1$.

In the pure majority-rule regime ($q=0$), the dynamics is symmetric under spin reversal, making $c=1/2$ a fixed point. Following a standard saddle-point treatment of the backward FP problem near the symmetric point (see Appendix~\ref{app:exit_q0}), one obtains an explicit scaling form for $E(c_0)$. The new element in the present setting is that the higher-order interaction structure enters this scaling form through a single parameter $\gamma(P)$, and we provide $\gamma(P)$ explicitly for the ensembles considered. In the central scaling regime, the exit probability takes a universal error-function form:
\begin{equation}
  E(c_0)\approx
  \frac{1}{2}\left[
    1 + \operatorname{erf}\left(
      \sqrt{\gamma(P)}
      \left(c_0-\tfrac{1}{2}\right)
    \right)\right],
  \label{eq:exit_prob_general}
\end{equation}
where $\gamma(P) = \mathcal{A}_1(P)/D_0(P)$. Here, $\mathcal{A}_1(P)$ is the average majority leverage defined in Eq.~\eqref{eq:A_1}, and $D_0(P)\equiv D(c=1/2;P)$ is the diffusion coefficient evaluated at the symmetric point for the hyperedge-size distribution $P(n)$. The explicit expressions for specific ensembles are provided in Eqs.~\eqref{eq:gamma_unif}--\eqref{eq:gamma_pl} of Appendix~\ref{app:exit_q0}. Equation~\eqref{eq:exit_prob_general} shows that, within the diffusion description, different higher-order structures are distinguished solely by $\gamma(P)$, which measures the ratio between majority amplification and stochastic fluctuations near the separatrix $c=1/2$.

The diffusion approximation underlying Eq.~\eqref{eq:exit_prob_general} assumes that the elementary increments $\Delta c$ are small, so that truncating the KM expansion at second order is accurate. This assumption holds well for the uniform and shifted-geometric ensembles, where hyperedge sizes are bounded or decay exponentially. For heavy-tailed power-law ensembles, however, sufficiently small exponents allow occasional hyperedges with $n=O(N)$, which induce macroscopic jumps in $c$ and violate the small-jump assumption. In that regime, the standard FP description tends to underestimate the steepness of the exit-probability curve. Accordingly, we use Eq.~\eqref{eq:exit_prob_general} as a scaling baseline and expect quantitative agreement primarily in parameter ranges where large-$n$ events are sufficiently suppressed (see Appendix~\ref{app:exit_q0} for details). A fully quantitative theory for strongly heavy-tailed cases would require the discrete master equation with nonlocal transitions, which is beyond the scope of the present work.

\begin{figure}[tb!]
    \centering
    \includegraphics[width=0.8\linewidth]{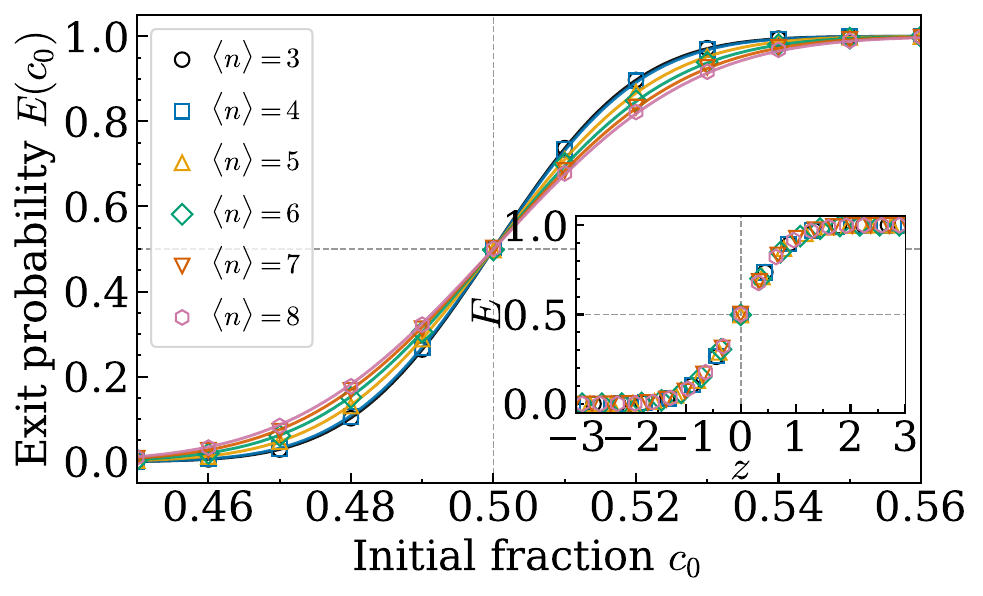}
    \caption{
    Exit probability $E$ as a function of the initial fraction $c_0$ in the pure majority-rule regime ($q=0$) for the shifted geometric hyperedge-size distribution. Results are shown for several mean hyperedge sizes $\langle n\rangle$ at $N=10^3$. Symbols denote MC simulation results averaged over $10^5$ independent realizations, while solid lines represent the diffusion-theory prediction given by Eq.~\eqref{eq:exit_prob_general}. The inset shows the corresponding scaling collapse when $E$ is plotted against the scaling variable $z=\sqrt{\gamma(P)}(c_0-1/2)$.
    }
    \label{fig:exit_prob}
\end{figure}

Nevertheless, Eq.~\eqref{eq:exit_prob_general} provides a robust scaling framework when the diffusion approximation applies. To make finite-size effects explicit, it is convenient to introduce the scaling variable $z = \sqrt{\gamma(P)}\left(c_0-1/2\right)$. In terms of $z$, Eq.~\eqref{eq:exit_prob_general} reads $  E(c_0)\simeq \frac{1}{2}\bigl[1+\operatorname{erf}(z)\bigr]$, so that all dependence on $N$ and on the hyperedge-size statistics is absorbed into the rescaling of the initial condition. Plotting $E$ as a function of $z$ collapses the data for different system sizes (at fixed $P$) onto a single universal curve, as illustrated in the inset of Fig.~\ref{fig:exit_prob}.

From the definition of $\gamma(P)$, the width of the transition region scales as $  \Delta c \sim \gamma(P)^{-1/2}
  = \sqrt{D_0(P)/\mathcal{A}_1(P)}$.  Since $D_0(P)\sim N^{-1}$ while $\mathcal{A}_1(P)$ is $O(1)$, we recover the standard MF finite-size scaling $\Delta c \sim N^{-1/2}$. Thus, for large but finite $N$, Eq.~\eqref{eq:exit_prob_general} quantifies how $P(n)$ controls the sharpness of the probabilistic crossover near $c_0=1/2$ through $\gamma(P)$, while in the thermodynamic limit the curve sharpens into a step function.

\section{Summary and Conclusion}

We have investigated a noisy majority-rule dynamics on annealed hypergraphs in which, at each update, a group size $n$ is first drawn from a prescribed distribution $P(n)$, after which a set of $n$ distinct agents is selected uniformly at random from the population. The selected hyperedge then either adopts the strict local majority with probability $1-q$ or undergoes a collective field-driven update with probability $q$, in which all selected agents adopt state $+1$ with probability $p$ and state $-1$ with probability $1-p$. Our main goal was to isolate how fluctuations in interaction-group size, encoded by $P(n)$, reshape ordering, disordering, and finite-size relaxation beyond the fixed-$n$ baseline.

At the up--down symmetric point $p=1/2$, the model exhibits an order--disorder transition controlled by the field-driven update probability $q$. By expanding the drift around the mixed fixed point, we derived a general expression for the symmetric critical threshold $q_c(P)$ in terms of the mean group size and a structural coefficient quantifying the average majority leverage. Analytical predictions for $n$-uniform, shifted geometric, and truncated power-law ensembles agree well with MC simulations. Physically, $q_c(P)$ quantifies the maximal level of field-driven updating compatible with macroscopic ordering; broader distributions that place more weight on large groups increase this tolerance by amplifying majority alignment. For $n$-uniform and geometric distributions, $q_c$ increases with the characteristic group size. For power-law ensembles, $q_c(\alpha)$ approaches the triplet limit $1/3$ as $\alpha \to \infty$, whereas heavy-tailed statistics at small $\alpha$ assign significant weight to large hyperedges and push $q_c$ toward unity.

In the pure majority-rule regime ($q=0$), we obtained logarithmic consensus-time scaling with a distribution-dependent prefactor. For initially balanced configurations, the prefactor contains contributions from both the escape from the unstable mixed state and the final approach to consensus. For unbalanced initial conditions, the mixed-state bottleneck is bypassed and the prefactor is controlled primarily by the statistics of the selected group sizes. In both cases, larger effective interaction sizes accelerate ordering. For truncated power-law ensembles, sufficiently heavy tails can suppress the logarithmic growth and lead to saturation or sublogarithmic relaxation under the natural cutoff.

In the disordered phase $q>q_c(P)$, we characterized the disordering time $T_{\mathrm{dis}}$, defined as the mean time required to reach the vicinity of the mixed state from a fully ordered initial configuration. Combining a diffusion approximation with a backward-FP treatment, we obtained the leading form $T_{\mathrm{dis}}\sim \mathcal{B}(P,q)\ln N$ when the corresponding logarithmic prefactor remains finite. For $n$-uniform and shifted geometric ensembles, this gives logarithmic growth throughout the disordered regime. For truncated power-law ensembles, the asymptotic behavior is more sensitive to the tail of $P(n)$: for $\alpha>5/2$, the logarithmic form persists, whereas broader distributions lead to stronger finite-size acceleration associated with rare large-group events. In the extreme-bias limits $p=0$ and $p=1$, where one consensus boundary remains absorbing, the consensus time follows the same distribution-controlled scaling as the unbalanced pure majority-rule case, highlighting the central role of $P(n)$ in setting ordering time scales.

We also analyzed the exit probability $E(c_0)$ in the pure majority-rule regime, i.e., the probability of reaching the all-$+1$ state from an initial fraction $c_0$ of $+1$ agents. Near the symmetric point, $E(c_0)$ collapses onto a universal error-function form governed by a single scaling variable combining $c_0$, the system size $N$, and the hyperedge statistics. Ensembles that favor larger groups exhibit a larger drift-to-diffusion ratio, thereby sharpening the crossover and making small initial imbalances more decisive. Conversely, ensembles dominated by small groups broaden the transition region and enhance the role of stochastic fluctuations. For sufficiently broad ensembles, rare large-group events can introduce quantitative corrections to the diffusion approximation, although the scaling framework still captures the qualitative sharpening mechanism.

In summary, our results show that the full group-size distribution $P(n)$ controls both the robustness of ordering through the symmetric threshold $q_c(P)$ and the characteristic relaxation time scales, including crossovers induced by rare large-group events. Natural extensions of this work include quenched hypergraphs, the interplay between higher-order and pairwise interactions, and time-dependent or competing external fields. These directions may help connect distribution-controlled higher-order ordering mechanisms to empirically grounded group-interaction data in social and technological systems.

\section*{Acknowledgments}
\textbf{R.~Muslim} was supported by the YST Program of the Asia Pacific Center for Theoretical Physics (APCTP), funded by the Science and Technology Promotion Fund and Lottery Fund of the Korean Government, and by the Management Talent Program of the National Research and Innovation Agency of Indonesia (BRIN). \textbf{J.-M.~Park} acknowledges support from an appointment to the JRG Program at APCTP through the Science and Technology Promotion Fund and Lottery Fund of the Korean Government. This work was also supported by the Korean Local Governments---Gyeongsangbuk-do Province and Pohang City. This work was supported by the National Research Foundation of Korea (NRF) grant funded by the Korea government (MSIT) (Grant No.~RS-2025-00557038). \textbf{R.~Anugraha~NQZ} acknowledges financial support from the Directorate of Research and Community Service, Ministry of Higher Education, Science, and Technology, through the Regular Fundamental Research Scheme under Contract Number 1262/UN1/DITLIT/Dit-Lit/PT.01.03/2026.

\bibliographystyle{apsrev4-2}

\appendix
\section{\label{app:rates}Rate Equations and Symmetric Criticality}

We derive the rate equations for the noisy majority-rule (MR) dynamics with field-driven group updates defined in Sec.~\ref{sec:model}. The configuration evolves through random sequential updates: in each elementary update, a group size $n\in\{3,\dots,n_{\max}\}$ is drawn from $P(n)$ and the corresponding active hyperedge is sampled according to the rule specified in Sec.~\ref{sec:model}. With probability $q$ (field-driven update), the selected group is assigned state $+1$ with probability $p$ and state $-1$ with probability $1-p$. With probability $1-q$ (majority rule), all selected agents adopt the within-group majority opinion; for even $n$, tie events ($\Sigma=0$) yield no change. Throughout, we measure time in Monte Carlo steps (MCS) using the rescaled variable $\tau=t/N$, where $t$ counts elementary updates, so that $\delta\tau=1/N$ per update.

For finite $N$, the exact composition of the selected hyperedge is hypergeometric. In the analytical treatment below, we use the annealed large-$N$ mean-field closure, in which the group composition is approximated by the binomial weight $B_{n,\ell}(c)=\binom{n}{\ell}c^\ell(1-c)^{n-\ell}$. This approximation becomes asymptotically exact in the well-mixed limit when $n/N\to0$ and allows compact expressions for the drift and the symmetric threshold.

Let $R(c)\equiv \langle N_{-\to+}\mid c\rangle$ and $L(c)\equiv \langle N_{+\to-}\mid c\rangle$ denote the conditional expectations of the numbers of $-1\to+1$ and $+1\to-1$ flips produced by a single elementary update, given the current density $c$ of $+1$ opinions. Within the annealed MF closure, these rates are approximated by
\begin{equation}
\begin{aligned}
  R(c) & \approx \sum_{n=3}^{n_{\max}} P(n)\left[(1-q)\,M_n(c)+q\,n\,p\,(1-c)\right],\\
  L(c) & \approx \sum_{n=3}^{n_{\max}} P(n)\left[(1-q)\,M_n(1-c)+q\,n\,(1-p)\,c\right],
\end{aligned}
\label{eq:app_rates}
\end{equation}
where $M_n(c)=\sum_{\ell=\lfloor n/2\rfloor+1}^{n} (n-\ell)\binom{n}{\ell}c^{\ell}(1-c)^{n-\ell}$ is the expected number of $-1\to+1$ flips generated by a majority update in a size-$n$ group when a strict $+1$ majority is present. For even $n$, ties do not contribute. The complementary term $M_n(1-c)$ accounts for strict $-1$ majorities.

Accordingly, the coarse-grained drift per unit $\tau$ is approximated as
\begin{align}
  v(c) &\equiv \frac{\langle \Delta c\rangle}{\delta\tau}
  = R(c)-L(c)
  \nonumber \\
  &\approx \sum_{n=3}^{n_{\max}} P(n)\Big\{q\,n(p-c)+(1-q)\Delta M_n(c)\Big\},
  \label{eq:app_drift}
\end{align}
with $\Delta M_n(c)\equiv M_n(c)-M_n(1-c)$. Solving the stationarity condition $v(c)=0$ for $q$ at arbitrary $c$ yields
\begin{equation}
  q(c) = \frac{\sum_{n=3}^{n_{\max}} P(n)\Delta M_n(c)}
           {\sum_{n=3}^{n_{\max}} P(n)\Delta M_n(c)+\big(c-p\big)\langle n\rangle},
  \label{eq:app_q_of_c}
\end{equation}
where $\langle n\rangle\equiv\sum_{n=3}^{n_{\max}} nP(n)$.

Equation~\eqref{eq:app_q_of_c} makes clear that, for $p\neq 1/2$, the stationary branch is generally shifted away from $c=1/2$. In particular, the balanced state $c=1/2$ is a fixed point only in the up--down symmetric case $p=1/2$. Therefore, the critical threshold derived below should be understood as the symmetric threshold at $p=1/2$, in agreement with the discussion in Sec.~\ref{sec:sub_section_iii_ii}.

For the up--down symmetric case $p=1/2$, the disordered state $c=1/2$ is always a fixed point. To analyze the order--disorder transition, we perform a linear expansion of the drift around this point. Let $c = 1/2 + \delta$, where $|\delta| \ll 1$.

Expanding the binomial probability weight $B_{n,\ell}(c) \equiv \binom{n}{\ell}c^{\ell}(1-c)^{n-\ell}$ to first order in $\delta$ yields
\begin{align}
  B_{n,\ell}(1/2+\delta) &= \binom{n}{\ell} \left(1/2+\delta\right)^{\ell} \left(1/2-\delta\right)^{n-\ell} \nonumber \\
  &\simeq 2^{-n}\binom{n}{\ell} \big[1 + 2\left(2\ell-n\right)\delta\big].
\end{align}
Substituting this into the majority-rule drift term $\Delta M_n(c) = M_n(c) - M_n(1-c)$ and using the antisymmetry around $c=1/2$, we obtain
\begin{align} \label{eq:DM_expand}
  \Delta M_n(c) &\simeq \sum_{\ell=\lfloor \frac{n}{2}\rfloor+1}^{n} \left(n-\ell\right) \left[ B_{n,\ell}\left(1/2+\delta\right) - B_{n,\ell}\left(1/2-\delta \right) \right] \nonumber \\
  &= \sum_{\ell=\lfloor \frac{n}{2}\rfloor+1}^{n} \left(n-\ell \right)\binom{n}{\ell} 2^{-n} \Big[ 4\left(2\ell-n \right)\delta \Big] \nonumber \\
  &= 2^{2-n} S_n \left(c-1/2\right),
\end{align}
where $S_n = \sum_{\ell=\lfloor n/2\rfloor+1}^{n} \left(n-\ell \right)\left(2\ell-n \right)\binom{n}{\ell}$.

The critical field-driven update probability follows from the condition $v'(c)|_{c=1/2}=0$. Differentiating Eq.~\eqref{eq:app_drift} with respect to $c$ and using Eq.~\eqref{eq:DM_expand}, we obtain
\begin{align}
   v'(1/2) = &  \left(1-q\right) \sum_{n=3}^{n_{\max}} P(n)\,2^{2-n} S_n -q\,\langle n\rangle .
\end{align}
Setting $v'(1/2)=0$ yields the symmetric critical threshold
\begin{align}
  q_c = \frac{\sum_{n=3}^{n_{\max}} P(n) 2^{2-n} S_n}
             {\sum_{n=3}^{n_{\max}} P(n) 2^{2-n} S_n + \langle n \rangle}
      = \frac{\mathcal{A}_1(P)}{\mathcal{A}_1(P)+\langle n \rangle},
  \label{eq:app_qc_general}
\end{align}
where $\mathcal{A}_1(P) \equiv \sum_{n=3}^{n_{\max}} P(n) \, 2^{2-n} S_n$ is the mean majority leverage.

\subsection{Annealed $n$-uniform distribution}

For the $n$-uniform ensemble with fixed group size $n$, we take
$P(n')=\delta_{n',n}$. Substituting this into Eq.~\eqref{eq:app_qc_general},
the sums collapse to single terms:
\begin{equation}
  q_c(n) = \frac{2^{2-n}S_n}{2^{2-n}S_n+n},
  \label{eq:app_qc_uniform_repeat}
\end{equation}
which coincides with Eq.~\eqref{eq:crit_q_uniform} in the main text.

\subsection{Geometric distribution}

For the truncated shifted geometric ensemble, we use
$P(n)=[(1-x)x^{n-3}]/[1-x^{n_{\max}-2}]$ for
$n=3,\dots,n_{\max}$ and $0<x<1$. The corresponding mean group size is
\begin{align}
  \langle n\rangle
  = &\sum_{n=3}^{n_{\max}} nP(n)
  \nonumber \\
  = & 3+\dfrac{x\left[1-(n_{\max}-2)x^{n_{\max}-3}+(n_{\max}-3)x^{n_{\max}-2}\right]}
  {(1-x)\left(1-x^{n_{\max}-2}\right)}.
  \label{eq:app_geo_mean}
\end{align}

The associated linear drift coefficient is
\begin{align}
  \mathcal{A}_1^{\mathrm{geo}}(x;n_{\max})
  &= \sum_{n=3}^{n_{\max}} P(n)\,2^{2-n}S_n \nonumber \\
  &= \frac{1-x}{1-x^{\,n_{\max}-2}}\sum_{n=3}^{n_{\max}} x^{n-3}\,2^{2-n}S_n.
  \label{eq:app_geo_Sn}
\end{align}

Substituting Eqs.~\eqref{eq:app_geo_mean} and \eqref{eq:app_geo_Sn} into
Eq.~\eqref{eq:app_qc_general} gives the finite-$n_{\max}$ critical threshold,
\begin{equation}
  q_c^{\mathrm{geo}}(x;n_{\max})
  =\frac{\mathcal{A}_1^{\mathrm{geo}}(x;n_{\max})}
         {\mathcal{A}_1^{\mathrm{geo}}(x;n_{\max})+\langle n\rangle}.
  \label{eq:app_qc_geo_finite}
\end{equation}

In the thermodynamic limit $n_{\max}\to\infty$, one obtains
$\langle n\rangle \to (3-2x)/(1-x)$ and
$\mathcal{A}_1^{\mathrm{geo}}(x;n_{\max}) \to
(1-x)\sum_{n=3}^{\infty} x^{n-3}2^{2-n}S_n$. Hence,
\begin{equation}
  q_c(x)
  =\frac{(1-x)^2\sum_{n=3}^{\infty} x^{n-3}2^{2-n}S_n}
         {(1-x)^2\sum_{n=3}^{\infty} x^{n-3}2^{2-n}S_n+(3-2x)},
  \label{eq:app_qc_geo_limit}
\end{equation}
which is the expression quoted in the main text [Eq.~\eqref{eq:crit_q_geo}].

\subsection{Power-law distribution}

For the truncated power-law ensemble, we consider $P(n)=n^{-\alpha}/Z_\alpha$ with
$Z_\alpha=\sum_{n=3}^{N} n^{-\alpha}$ and $\alpha>0$.
Note that varying $\alpha$ changes not only the tail weight, but also the truncated mean hyperedge size
\begin{equation}
  \langle n\rangle_{N}(\alpha)\equiv \sum_{n=3}^{N} n\,P(n)
  =\frac{\sum_{n=3}^{N} n^{1-\alpha}}{\sum_{n=3}^{N} n^{-\alpha}}
  =\frac{B_N(\alpha)}{Z_\alpha},
\label{eq:mean_n_powerlaw}
\end{equation}
which will be useful when interpreting $\alpha$-dependent trends.
Returning to Eq.~\eqref{eq:app_qc_general}, it is convenient to rewrite the finite-$N$ critical point as
\begin{equation}
  q_c(\alpha;N)=\frac{A_N(\alpha)}{A_N(\alpha)+B_N(\alpha)},
  \label{eq:qc_pl_basic}
\end{equation}
where $A_N(\alpha) \equiv \sum_{n=3}^{N} n^{-\alpha}2^{2-n}S_n$ and
$B_N(\alpha) \equiv \sum_{n=3}^{N} n^{1-\alpha}$.
In Eq.~\eqref{eq:qc_pl_basic}, $B_N(\alpha)$ is directly proportional to the first moment of the size distribution [cf.~Eq.~\eqref{eq:mean_n_powerlaw}], while $A_N(\alpha)$ encodes the majority-amplification factor $2^{2-n}S_n$ and therefore depends on the full distribution of group sizes, not only on $\langle n\rangle_N$.

To estimate $A_N(\alpha)$ for large $n$, we use a Gaussian approximation for the binomial coefficient,
\begin{equation}
  \binom{n}{\ell}\approx 2^n\sqrt{\frac{2}{\pi n}}\,
  \exp\left[-\frac{2 \left(\ell-n/2 \right)^2}{n}\right].
\end{equation}
Writing $\ell=n/2+j$ and approximating the sum by an integral, one obtains the leading asymptotic behavior
\begin{equation}
  2^{2-n}S_n \sim \sqrt{\frac{2}{\pi}}\,n^{3/2}\qquad \left(n\gg 1 \right).
  \label{eq:Sn_asym}
\end{equation}
Substituting this into the definition of $A_N(\alpha)$, we find
\begin{equation}
  A_N(\alpha)
  \propto \sum_{n=3}^{N} n^{3/2-\alpha}.
  \label{eq:AN_asym}
\end{equation}
From Eq.~\eqref{eq:AN_asym}, one finds that $A_N(\alpha)$ scales as $N^{5/2-\alpha}$ for $\alpha<5/2$, grows logarithmically as $\ln N$ for $\alpha=5/2$, and converges to a finite constant
$A(\alpha)=\sum_{n=3}^{\infty} n^{-\alpha}2^{2-n}S_n$ for $\alpha>5/2$.

For the moment term $B_N(\alpha)=\sum_{n=3}^{N} n^{1-\alpha}$, standard estimates give
$B_N(\alpha)\propto N^{2-\alpha}$ for $\alpha<2$, $B_N(2)\sim\ln N$, and
$B_N(\alpha)\to \zeta(\alpha-1,3)$ for $\alpha>2$, where $\zeta(a,b)$ is the Hurwitz zeta function.
Consequently, for the natural cutoff $n_{\max}=N$, the mean size $\langle n\rangle_N(\alpha)=B_N(\alpha)/Z_\alpha$ becomes $N$ dependent for $\alpha\le 2$ and diverges as $N\to\infty$, reflecting the increasing weight of large hyperedges as $\alpha$ decreases.

Combining these asymptotics with Eq.~\eqref{eq:qc_pl_basic}, we obtain
\begin{equation}
  q_c(\alpha) \equiv \lim_{N\to\infty} q_c(\alpha; N) =
  \begin{cases}
    1, & \alpha\le 5/2,\\[8pt]
    \displaystyle \frac{A(\alpha)}{A(\alpha)+\zeta(\alpha-1,3)},
    & \alpha> 5/2,
  \end{cases}
  \label{eq:qc_pl_regimes_final}
\end{equation}
where $A(\alpha)=\sum_{n=3}^{\infty} n^{-\alpha}2^{2-n}S_n$.
Thus, under the natural cutoff, the symmetric critical threshold tends to unity for all $\alpha\le 5/2$ and converges to a finite value in $(0,1)$ for $\alpha>5/2$.

In the limit $\alpha\to\infty$, the distribution concentrates on the smallest hyperedge size, $P(n)\to\delta_{n,3}$.
For $n=3$, one has $S_3=\sum_{\ell=2}^{3} (3-\ell)(2\ell-3)\binom{3}{\ell}=3$, so that $2^{2-3}S_3=3/2$. Using Eq.~\eqref{eq:app_qc_general} specialized to the $n=3$ case, we recover
\begin{equation}
  q_c(\infty)=\frac{3/2}{3/2+3}=\frac13,
\end{equation}
in agreement with the main-text result.

\section{\label{appendixB}Consensus Time}
\subsection{Pure majority-rule dynamics with an initially balanced state}

In this appendix, we restrict attention to the pure majority-rule regime $q=0$, for which both fully ordered states $c=0$ and $c=1$ are absorbing. Accordingly, the consensus time $T_{\mathrm{cons}}$ is well defined as the mean first-passage time to one of these two absorbing boundaries. In the absence of the field-driven branch, the dynamics is up--down symmetric, and it is sufficient to analyze one side of the flow, e.g., from the vicinity of the unstable fixed point $c=1/2$ toward the absorbing boundary at $c=0$; the opposite branch follows by symmetry.

In the MF limit, the diffusion term is negligible since $D(c)=\mathcal{O}(1/N)$. The consensus time is obtained by integrating the inverse drift characteristic $d\tau = dc/v(c)$ from the backward Kolmogorov equation:
\begin{equation}
  \label{eq:relax_int}
  T \approx \int_{c_f}^{c_i}-\frac{dc}{v(c)}.
\end{equation}
For the lower branch, the integration bounds are fixed by the microscopic fluctuation scales: the initial state is taken just outside the unstable fixed point, $c_i=1/2-N^{-1/2}$, and the final state is set at the absorption scale, $c_f=1/N$.

The integral in Eq.~\eqref{eq:relax_int} is dominated by the divergences near the two fixed points $c=0$ and $c=1/2$. We can therefore approximate the total time as the sum of the transit times across these two bottlenecks:
\begin{equation}
  T \approx T_{\text{abs}} + T_{\text{cen}}.
\end{equation}
For a general hyperedge-size distribution $P(n)$, the drift linearizes near these fixed points as follows:

Near the absorbing boundary ($c\to 0$), the leading contribution to the drift comes from sampled groups containing a single $+1$ agent. For a group of size $n$, such configurations occur with probability $n c + O(c^2)$; under strict majority rule, that lone $+1$ agent flips to $-1$, yielding one downward flip. Configurations containing two or more $+1$ agents are $O(c^2)$ and thus subleading. Averaging over $P(n)$ then gives the linear boundary-layer drift $v(c)\simeq -\sum_n P(n)\,n\,c=-\langle n\rangle c$. Introducing an $N$-independent cutoff $k_\ast=O(1)$ that separates the boundary layer from the bulk, we obtain
\begin{equation}
  T_{\mathrm{abs}} \approx \int_{1/N}^{k_\ast}\frac{dc}{\langle n\rangle c}
  = \frac{1}{\langle n\rangle}\ln N + O(1).
\end{equation}
The precise value of $k_\ast$ affects only the additive constant and does not modify the leading logarithmic term.

Near the unstable fixed point $c=1/2$, the drift behaves as $v(c)\simeq \mathcal{A}_1(P)(c-1/2)$. Introducing an $N$-independent cutoff $\delta_\ast=O(1)$ for the linear neighborhood of $c=1/2$, we find
\begin{align}
  T_{\mathrm{cen}} & \approx \int_{1/2-\delta_\ast}^{\,1/2-1/\sqrt{N}}
  -\frac{dc}{\mathcal{A}_1(P)(c-1/2)} \nonumber \\
  &= \frac{1}{2\mathcal{A}_1(P)}\ln N + O(1).
\end{align}
As above, the precise value of $\delta_\ast$ contributes only to the subleading constant and does not affect the coefficient of $\ln N$.

Combining these, the general scaling for the consensus time is
\begin{equation}
  \label{eq:T_general_formula}
  T(N) \sim \left[ \frac{1}{\langle n \rangle} + \frac{1}{2\mathcal{A}_1(P)} \right] \ln N.
\end{equation}
We now evaluate the specific coefficients $\langle n \rangle$ and $\mathcal{A}_1(P)$ for the different hyperedge-size ensembles.

\subsubsection{\label{appendix:B1} Annealed $n$-uniform distribution}
For the $n$-uniform ensemble, the coefficients are simply $\langle n \rangle = n$ and $\mathcal{A}_1(n) = 2^{2-n}S_n$. Substituting these into Eq.~\eqref{eq:T_general_formula} immediately yields
\begin{equation}
  T(n,N) \sim \left[\frac{1}{n} + \frac{2^{n-3}}{S_n}\right]\ln N,
\end{equation}
recovering Eq.~\eqref{eq:prefac_uniform_main} in the main text.

\subsubsection{Geometric distribution}
For the shifted geometric distribution, the mean size is $\langle n \rangle = (3-2x)/(1-x)$,
while the majority-leverage term $\mathcal{A}_1(P)$ is given by Eq.~\eqref{eq:app_geo_Sn}.
Substituting these coefficients into Eq.~\eqref{eq:T_general_formula} yields
\begin{equation}
  T(N,x)\sim
  \left[
    \frac{1-x}{3-2x}
    +
    \frac{1}{(1-x)\sum_{n\ge 3} x^{n-3}2^{3-n}S_n}
  \right]\ln N,
\end{equation}
which matches Eq.~\eqref{eq:prefac_geo_main} in the main text.

\subsubsection{\label{sub:subsub_pl}Power-law distribution}

We consider an annealed power-law ensemble where hyperedge sizes $n \in \{3, \dots, N\}$ are drawn from
\begin{equation}
  P_N(n)=\frac{n^{-\alpha}}{Z_N(\alpha)},\qquad
  Z_N(\alpha)=\sum_{n=3}^{N} n^{-\alpha},
\end{equation}
with exponent $\alpha>0$.
To apply the general consensus-time formula in Eq.~\eqref{eq:T_general_formula}, we identify the two relevant macroscopic rates: the absorbing-boundary contribution is governed by the first moment of the size distribution, whereas the central contribution is governed by the majority-leverage sum.
Accordingly,
\begin{align}
  \langle n \rangle &= \frac{B_N(\alpha)}{Z_N(\alpha)},
  \qquad
  B_N(\alpha) = \sum_{n=3}^{N} n^{1-\alpha}, \\
  2\mathcal{A}_1(P) &= \frac{C_N(\alpha)}{Z_N(\alpha)},
 \qquad
  C_N(\alpha) = \sum_{n=3}^{N} n^{-\alpha} 2^{3-n} S_n .
\end{align}
Substituting these into Eq.~\eqref{eq:T_general_formula} yields the prefactor
\begin{equation}
  T(N,\alpha)\sim
  \left[
    \frac{Z_N(\alpha)}{B_N(\alpha)}
    +\frac{Z_N(\alpha)}{C_N(\alpha)}
  \right]\ln N.
  \label{eq:pl_master}
\end{equation}

To analyze the thermodynamic limit, we determine the asymptotic behavior of the three sums.
The normalization $Z_N(\alpha)$ and the first moment $B_N(\alpha)$ follow standard power-law sum rules:
\begin{widetext}
\begin{equation}\label{eq:ZA_1}
  Z_N(\alpha) \sim
  \begin{cases}
    \dfrac{N^{1-\alpha}}{1-\alpha}, & \alpha<1,\\[6pt]
    \ln N, & \alpha=1,\\[6pt]
    \zeta(\alpha,3), & \alpha>1,
  \end{cases}
  \qquad
  B_N(\alpha) \sim
  \begin{cases}
    \dfrac{N^{2-\alpha}}{2-\alpha}, & \alpha<2,\\[6pt]
    \ln N, & \alpha=2,\\[6pt]
    \zeta(\alpha-1,3), & \alpha>2.
  \end{cases}
\end{equation}
\end{widetext}

For the majority-leverage sum $C_N(\alpha)$, we use the large-$n$ estimate
$S_n \sim \sqrt{2/\pi}\,n^{3/2}2^{n-2}$,
which implies $n^{-\alpha}2^{3-n}S_n \sim 2\sqrt{2/\pi}\,n^{3/2-\alpha}$.
Hence the divergence threshold is $\alpha=5/2$. Accordingly,
\begin{equation}\label{eq:A2}
  C_N(\alpha)\sim
  \begin{cases}
    N^{5/2-\alpha}, & \alpha<5/2,\\[6pt]
    \ln N, & \alpha=5/2,
  \end{cases}
\end{equation}
whereas for $\alpha>5/2$ the sum converges to the finite limit, $C_\infty(\alpha)=\sum_{n=3}^{\infty} n^{-\alpha}2^{3-n}S_n$.

Based on these asymptotics, we identify three distinct scaling regimes for the consensus-time prefactor $\mathcal{B}(\alpha)$:

\begin{enumerate}
    \item Regime $0<\alpha\le 2$: The boundary contribution $Z_N(\alpha)/B_N(\alpha)$ vanishes as $N\to\infty$, scaling as $N^{-1}$ for $0<\alpha<1$, as $(\ln N)/N$ for $\alpha=1$, as $N^{\alpha-2}$ for $1<\alpha<2$, and as $(\ln N)^{-1}$ for $\alpha=2$. In the same interval, $C_N(\alpha)$ diverges, so the central contribution also vanishes: $Z_N(\alpha)/C_N(\alpha)\sim N^{-3/2}$ for $0<\alpha<1$, $(\ln N)/N^{3/2}$ for $\alpha=1$, $N^{\alpha-5/2}$ for $1<\alpha<2$, and $N^{-1/2}$ for $\alpha=2$. Hence $\mathcal{B}(\alpha)\to 0$, implying that the consensus time grows more slowly than $\ln N$ and may saturate in the large-$N$ limit.

    \item Regime $2 < \alpha \le 5/2$: The sums $Z_N(\alpha)$ and $B_N(\alpha)$ converge, whereas $C_N(\alpha)$ still diverges (or grows logarithmically at $\alpha=5/2$). Consequently, the central contribution $Z_N(\alpha)/C_N(\alpha)$ vanishes. The dynamics is then controlled solely by the absorbing-boundary bottleneck:
    \begin{equation}
      \mathcal{B}(\alpha) = \frac{\zeta(\alpha,3)}{\zeta(\alpha-1,3)}.
    \end{equation}

    \item Regime $\alpha > 5/2$: All three sums converge. Both the absorbing boundary and the unstable central point contribute to the leading time scale, yielding
    \begin{equation}
      \mathcal{B}(\alpha) =
      \frac{\zeta(\alpha,3)}{\zeta(\alpha-1,3)}
      + \frac{\zeta(\alpha,3)}{C_\infty(\alpha)}.
    \end{equation}
\end{enumerate}

Summarizing these results, the consensus-time prefactor $\mathcal{B}(\alpha)$ is given by
\begin{equation}\label{eq:B_pl_alpha}
  \mathcal{B}(\alpha)=
  \begin{cases}
    0, & \alpha \le 2,\\[6pt]
    \dfrac{\zeta(\alpha,3)}{\zeta(\alpha-1,3)}, & 2<\alpha\leq\dfrac{5}{2},\\[12pt]
    \dfrac{\zeta(\alpha,3)}{\zeta(\alpha-1,3)}
    +\dfrac{\zeta(\alpha,3)}{\sum_{n=3}^{\infty} n^{-\alpha}2^{3-n}S_n},
    & \alpha>\dfrac{5}{2},
  \end{cases}
\end{equation}
consistent with Eq.~\eqref{eq:B_pl_limits} in the main text.

\begin{figure}[t]
    \centering
    \includegraphics[width=0.6\linewidth]{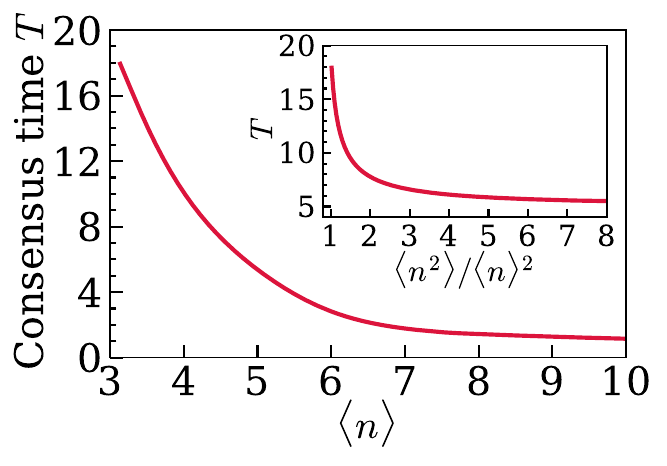}
    \caption{Consensus time $T_{\mathrm{cons}}$ for the power-law hyperedge-size distribution, reparametrized in terms of the moments of $P(n)$. The main panel shows $T_{\mathrm{cons}}$ versus the mean hyperedge size $\langle n\rangle$, while the inset shows the same result versus the fluctuation ratio $\langle n^2\rangle/\langle n\rangle^2$. In both cases, $T_{\mathrm{cons}}$ is evaluated using Eq.~\eqref{eq:B_pl_alpha} with $N=10^5$. The inset is restricted to $\alpha>3$, where the second moment is finite.}
    \label{fig:app_powerlaw_moments}
\end{figure}

For completeness, Fig.~\ref{fig:app_powerlaw_moments} re-expresses the power-law result in terms of the moments of $P(n)$. For $\alpha>2$, the mean hyperedge size is
$\langle n\rangle=\zeta(\alpha-1,3)/\zeta(\alpha,3)$, so that
$T_{\mathrm{cons}}\sim \mathcal{B}(\alpha)\ln N$ may be viewed as a function of $\langle n\rangle$, as shown in the main panel. For $\alpha>3$, where the second moment is finite, the same result may also be parametrized by the fluctuation ratio
$\langle n^2\rangle/\langle n\rangle^2=\zeta(\alpha-2,3)\zeta(\alpha,3)/\zeta(\alpha-1,3)^2$, as shown in the inset.

\section{\label{appendixC}Consensus Time in the Field-driven Regime under Extreme Bias}

In this appendix, we consider the field-driven regime $q>0$ under extreme bias, $p\in\{0,1\}$. In these two cases, one of the two fully ordered states remains absorbing: $c=1$ for $p=1$ and $c=0$ for $p=0$. This is the only parameter range with $q>0$ in which a strict consensus time $T_{\mathrm{cons}}$ remains well defined, in agreement with the discussion in Sec.~\ref{sec:sub_section_iv_ii}.

For $p=1$, both update branches favor the same boundary state $c=1$: in the majority-rule branch, any selected group with a strict $+1$ majority eliminates minority agents, while in the field-driven branch the whole selected group is set to $+1$ with probability one. The case $p=0$ is fully symmetric, with the absorbing state shifted to $c=0$. Thus, in the extreme-bias limit, the majority and field-driven mechanisms act constructively toward the same absorbing boundary.

Because of this constructive alignment, the drift does not vanish at $c=1/2$, and the nearly balanced bottleneck associated with the symmetric case is absent. Consequently, the consensus time is controlled solely by the relaxation of the boundary layer near the absorbing state. For initial conditions bounded away from the separatrix, i.e., with fixed $c_0\neq 1/2$ as $N\to\infty$, the dynamics is drift dominated and the mean first-passage time is obtained from the deterministic time-of-flight integral in Eq.~\eqref{eq:relax_int} where $c_i$ is an $O(1)$ initial condition in the basin of the absorbing state and $c_f$ is the microscopic cutoff near the absorbing boundary.

In the boundary layer, the drift linearizes as $v(c)\simeq \langle n\rangle(1-c)$ for $p = 1$, and, by symmetry, $v(c)\simeq -\langle n\rangle c$ for $p = 0$,
where $\langle n\rangle=\sum_n nP(n)$ is the mean hyperedge size. Equivalently, in the two extreme-bias cases the boundary-layer drift can be written compactly as $v(c)\simeq \langle n\rangle (p-c)$ for $p\in\{0,1\}$.

Integrating Eq.~\eqref{eq:relax_int} from an $O(1)$ initial condition to the absorbing cutoff $c_f=1-1/N$ for $p=1$ (or $c_f=1/N$ for $p=0$) gives
\begin{equation}
  T_{\mathrm{cons}} \sim \frac{1}{\langle n\rangle}\ln N.
\end{equation}
Thus, the leading asymptotic behavior coincides with that of the pure majority-rule dynamics started from an unbalanced initial condition: the central bottleneck contributes no logarithmic term, and the entire consensus time is set by the final elimination of the minority near the absorbing boundary.

Specializing this result to the three hyperedge-size ensembles immediately reproduces the prefactors reported in Eqs.~\eqref{eq:unif_cons_noiseless}--\eqref{eq:pl_cons_noiseless} of the main text. Therefore, in the extreme-bias field-driven regime, the consensus-time prefactor is independent of $q$ at leading order, because both update branches generate the same linear restoring drift toward the absorbing state.

\section{\label{app:disordering_time}Disordering Time in the Symmetric Case}

In this appendix, we restrict attention to the up--down symmetric case $p=1/2$. Only in this case is the mixed state centered at $c=1/2$, and only in this case is the threshold $q_c$ derived in Appendix~\ref{app:rates} the relevant critical point controlling the stability of that symmetric mixed state. For $p\neq 1/2$, the attractor is shifted away from $c=1/2$, so the long-time behavior is more appropriately described in terms of relaxation toward a biased attractor rather than by the disordering time defined here.

In the disordered phase $q>q_c$, the mixed state $c=1/2$ is linearly stable. For large $N$, the diffusion term is subleading, so the entrance into the disordered basin is governed by the drift $v(c)$. Using Eq.~\eqref{eq:landau_expansion} from the main text, the linearized drift at $p=1/2$ is $v(c)\simeq -\lambda(c-1/2)$, with $\lambda \equiv q\langle n\rangle-(1-q)\mathcal{A}_1(P)>0$ in the disordered phase.

For an initial condition fixed away from $c=1/2$, integrating down to the fluctuation boundary $c_f=1/2+1/\sqrt{N}$ gives
\begin{align}\label{eq:T_general_app}
  T(N,q,P) &\approx \int_{1}^{1/2+N^{-1/2}} \frac{dc}{v(c)} \nonumber \\
  &\sim \frac{\ln N}{2\big[q\langle n\rangle-(1-q)\mathcal{A}_1(P)\big]}.
\end{align}
Thus, the logarithmic growth arises from the final approach to the stable mixed state, whose fluctuation width is $O(N^{-1/2})$. The prefactor diverges as the relaxation rate tends to zero, signaling critical slowing down at the symmetric threshold.

To parametrize distances from criticality, write $q=\kappa q_c$ with $\kappa>1$. Here $q_c$ refers exclusively to the symmetric threshold at $p=1/2$. Substituting $q_c$ from Eq.~\eqref{eq:app_qc_general} into the denominator of Eq.~\eqref{eq:T_general_app} simplifies the bracketed term to $(\kappa-1)\mathcal{A}_1(P)$. Consequently, Eq.~\eqref{eq:T_general_app} becomes
\begin{equation}
  T(N,\kappa,P)\sim \frac{\ln N}{2(\kappa-1)\mathcal{A}_1(P)},
\end{equation}
showing that $T$ scales linearly with $\ln N$ and diverges as $(\kappa-1)^{-1}$ near criticality.

\subsection{Annealed $n$-uniform distribution}

For the annealed $n$-uniform ensemble one has $\langle n\rangle=n$ and $\mathcal{A}_1(n)=2^{2-n}S_n$. Substituting these into Eq.~\eqref{eq:T_general_app} gives
\begin{equation}
  T(N,q,n)\sim\frac{\ln N}{2\big[q\,n-(1-q)2^{2-n}S_n\big]},
\end{equation}
in agreement with Eq.~\eqref{eq:polar_time_unif} in the main text.
Equivalently, in terms of $\kappa>1$,
\begin{align}
  T(N,\kappa,n)
  \sim \frac{\ln N}{2(\kappa-1)\mathcal{A}_1(n)} = \frac{2^{n-3}}{(\kappa-1)S_n}\ln N.
\end{align}

\subsection{Annealed geometric distribution}

For the annealed geometric distribution, substituting $\langle n\rangle$ and $\mathcal{A}_1(x)$ from Eqs.~\eqref{eq:app_geo_mean} and \eqref{eq:app_geo_Sn} into Eq.~\eqref{eq:T_general_app} gives
\begin{equation}
  T(N,q,x)\sim
  \frac{\ln N}{
    2\left[q\dfrac{3-2x}{1-x}
    -(1-q)\mathcal{A}_1(x)\right]
  },
\end{equation}
in agreement with Eq.~\eqref{eq:polar_time_geo} in the main text. Equivalently, in terms of the dimensionless distance from the symmetric threshold $\kappa>1$,
\begin{align}
  T(N,\kappa,x)
  &\sim \frac{\ln N}{2(\kappa-1)\mathcal{A}_1(x)} \nonumber\\
  &= \dfrac{\ln N}{(\kappa-1)(1-x)\displaystyle\sum_{n\ge3}x^{n-3}2^{3-n}S_n}.
  \label{eq:disordering_time_geo_kappa}
\end{align}

\subsection{Truncated power-law distribution}
Consider the annealed power-law ensemble $P(n)=n^{-\alpha}/Z_N(\alpha)$ with a hard cutoff $n_{\max}=N$ and exponent $\alpha>0$. Using
\begin{equation}
  \langle n\rangle=\frac{B_N(\alpha)}{Z_N(\alpha)},
  \qquad
  \mathcal{A}_1(\alpha)=\frac{A_N(\alpha)}{Z_N(\alpha)},
\end{equation}
with
\begin{equation}
  A_N(\alpha)=\sum_{n=3}^{N}n^{-\alpha}2^{2-n}S_n,
  \qquad
  B_N(\alpha)=\sum_{n=3}^{N}n^{1-\alpha},
\end{equation}
Eq.~\eqref{eq:T_general_app} yields
\begin{equation}
  \mathcal{B}(q,\alpha;N)
  = \frac{Z_N(\alpha)}
         {2q\,B_N(\alpha)-2(1-q)A_N(\alpha)}.
  \label{eq:pl_pref_general}
\end{equation}

The disordering-time analysis requires the linear stability condition $\lambda_N>0$, where $\lambda_N \equiv q\langle n\rangle_N-(1-q)\mathcal{A}_{1,N}(\alpha)$. Equivalently,
\begin{equation}
  2q\,B_N(\alpha)-2(1-q)A_N(\alpha) > 0,
  \label{eq:pl_stability_cond}
\end{equation}
which is identical to $q>q_c(N)$ with
\begin{equation}
  q_c(N)=\frac{A_N(\alpha)}{A_N(\alpha)+B_N(\alpha)}.
  \label{eq:qcN_pl}
\end{equation}

For $\alpha\le 5/2$, $A_N(\alpha)$ diverges with $N$ while $B_N(\alpha)$ grows more slowly (or converges), implying $q_c(N)\to 1$. Therefore, at fixed absolute $q<1$ the inequality \eqref{eq:pl_stability_cond} eventually fails and the mixed fixed point becomes unstable; in that case the MFPT to reach the neighborhood of $c=1/2$ is not the relevant large-$N$ descriptor. In finite systems one may still enforce the disordered regime by choosing $q>q_c(N)$, and then Eqs.~\eqref{eq:T_general_app} and \eqref{eq:pl_pref_general} remain valid with the $N$-dependent prefactor $\mathcal{B}(q,\alpha;N)$.

For $\alpha > 5/2$, all three sums converge to finite constants: $  Z_N(\alpha)\to \zeta(\alpha,3),\,
  B_N(\alpha)\to \zeta(\alpha-1,3),\,
  A_N(\alpha)\to A(\alpha)\equiv\sum_{n=3}^{\infty} n^{-\alpha}2^{2-n}S_n$. 
The prefactor therefore becomes independent of $N$:
\begin{equation}\label{eq:prefactor_dis_alpha}
  \mathcal{B}(q,\alpha)
  = \frac{\zeta(\alpha,3)}
         {2q\,\zeta(\alpha-1,3)-2(1-q)A(\alpha)}.
\end{equation}
Using the parameterization $q=\kappa q_c(\alpha)$ with $\kappa>1$ (well defined here only in the symmetric case and only when $q_c(\alpha)<1$, i.e., for $\alpha>5/2$), Eq.~\eqref{eq:prefactor_dis_alpha} simplifies to
\begin{equation}
  \mathcal{B}(\kappa,\alpha)
  = \frac{\zeta(\alpha,3)}{2(\kappa-1)A(\alpha)}.
\end{equation}
Thus, we recover the logarithmic scaling $  T_{\mathrm{dis}} \sim \mathcal{B}(\kappa,\alpha)\ln N$ in this regime.

These results correspond directly to the main-text discussion around Eqs.~\eqref{eq:lambdaN_pl}--\eqref{eq:Tdis_pl_general} and the scaling form \eqref{eq:Tdis_kappa}.

\section{Exit Probability for $q=0$}
\label{app:exit_q0}

In the pure majority-rule regime ($q=0$), the dynamics is up-down symmetric and
$c=1/2$ is the symmetry point. The exit probability $E(c)$ satisfies the backward
FP equation in Eq.~\eqref{eq:FP_exit_main}, with boundary conditions $E(0)=0$
and $E(1)=1$. Its standard quadrature form is
\begin{equation}
  E(c_0)=
  \frac{\displaystyle \int_{0}^{c_0}\exp[-\Phi(y)]\,dy}
       {\displaystyle \int_{0}^{1}\exp[-\Phi(y)]\,dy},
  \label{eq:app_exit_quadrature}
\end{equation}
where $\Phi(y)\equiv \int_{1/2}^{y} [2v(u)/D(u)]\,du.$
The lower limit $1/2$ is chosen for convenience, since the expansion is carried out around the symmetric point; changing this reference only adds a constant to $\Phi(y)$, which cancels between numerator and denominator in Eq.~\eqref{eq:app_exit_quadrature}.

Near $c=1/2$, we approximate the transport coefficients by their leading forms. Using
Eq.~\eqref{eq:D_from_RL_exact} with $v(1/2)=0$ and Eqs.~\eqref{eq:second_moment_decomp}-\eqref{eq:SnMR_explicit} at $q=0$, we obtain
\begin{equation}
  D_0(P)\equiv D(1/2)=\frac{1}{N}\sum_{n\ge 3}P(n)\,K_n,
  \label{eq:D_app}
\end{equation}
where
\begin{align}
  K_n \equiv & S_n^{\mathrm{MR}}(1/2)
  \nonumber \\
  =&2^{-n}\!\left[
  \sum_{\ell>n/2}(n-\ell)^2\binom{n}{\ell}
  +\sum_{\ell<n/2}\ell^2\binom{n}{\ell}
  \right].
\end{align}
The drift is linearized as $v(c)\simeq \mathcal{A}_1(P)\left(c-1/2\right)$, so that
\begin{align}
  \Phi(y)
  &\simeq \frac{2\mathcal{A}_1(P)}{D_0(P)}
  \int_{1/2}^{y}\left(u-1/2\right)\,du = \gamma(P)\left(y-1/2\right)^2,
\end{align}
where $\gamma(P)\equiv \mathcal{A}_1(P)/D_0(P)$.

Substituting this harmonic form into Eq.~\eqref{eq:app_exit_quadrature} and changing variables to $s=y-1/2$, we obtain
\begin{equation}
  E(c_0)\simeq
  \dfrac{\displaystyle\int_{-1/2}^{\,c_0-1/2}\exp{\left[-\gamma(P)s^2 \right]}\,ds}
       {\displaystyle\int_{-1/2}^{\,1/2} \exp{\left[-\gamma(P)s^2 \right]}\,ds}.
  \label{eq:app_exit_finite_window}
\end{equation}
For large $N$, one has $\gamma(P)\propto N\gg 1$, so the Gaussian kernel is localized near $s=0$ with width $O(N^{-1/2})$. The finite bounds $\pm 1/2$ may therefore be extended to $\pm\infty$ with exponentially small corrections, yielding
\begin{align}
  E(c_0)
  &\approx \dfrac{ \displaystyle \int_{-\infty}^{c_0-1/2} \exp\left[-\gamma(P)s^2\right] ds}
  { \displaystyle\int_{-\infty}^{+\infty} \exp\left[-\gamma(P)s^2\right] ds} \nonumber \\
  & =\frac{1}{2}\left[
    1 + \operatorname{erf}\left(
      \sqrt{\gamma(P)}\,
      \left(c_0-1/2\right)
    \right)\right].
  \label{eq:app_exit_general}
\end{align}
In the second line we used the standard Gaussian integral and its representation in terms of the error function.

The explicit form of $\gamma(P)$ for the $n$-uniform ensemble is
\begin{equation}\label{eq:gamma_unif}
\gamma_{\mathrm{u}}(n) = N\,\frac{2^{2-n}S_{n}}{K_{n}}.
\end{equation}
For the shifted geometric distribution,
\begin{equation}\label{eq:gamma_geo}
  \gamma_{\mathrm{g}}(x)
  = N \frac{\sum_{n=3}^{\infty} x^{n-3} 2^{2-n} S_n}
           {\sum_{n=3}^{\infty} x^{n-3} K_n}.
\end{equation}
Finally, for the truncated power-law distribution,
\begin{equation} \label{eq:gamma_pl}
  \gamma_{\mathrm{pl}}(N,\alpha)
  = N \frac{\sum_{n=3}^{N} n^{-\alpha} 2^{2-n} S_n}
           {\sum_{n=3}^{N} n^{-\alpha} K_n}.
\end{equation}

\section{Monte Carlo Simulation Protocol}
\label{app:MC_protocol}

The MC simulations follow the microscopic update rule defined in
Sec.~\ref{sec:model}. At each elementary update, a group size $n$ is first drawn from the prescribed distribution $P(n)$, and then $n$ distinct agents are sampled uniformly at random from the population without replacement. The resulting active hyperedge is therefore a set of distinct agents. After computing the pre-update sum
$\Sigma=\sum_{j=1}^{n}\sigma_{i_j}$, the selected agents are updated according to the majority or field-driven rule. One MCS consists of $N$
such elementary updates, and time is reported in the rescaled unit $\tau=t/N$.

Unless stated otherwise, the initial condition is prepared by assigning state
$+1$ independently with probability $c_0$ and state $-1$ with probability
$1-c_0$. Thus, $c_0=1/2$ corresponds to a balanced initial condition in
expectation, while $c_0\in\{0,1\}$ gives fully ordered initial states. For the exit probability $E(c_0)$, the initial fraction is fixed by preparing $N_+=c_0N$ agents in state $+1$, with $c_0N$ chosen to be an integer, and each realization is evolved until it reaches either $c=1$ or $c=0$.

For the consensus time $T_{\mathrm{cons}}$, each realization is evolved until it first reaches an absorbing consensus state, $c=0$ or $c=1$. This applies directly in the pure majority-rule regime ($q=0$), and for $q>0$ only in the extreme-bias cases $p=0$ and $p=1$, where absorbing consensus remains well defined.

For the disordering time $T_{\mathrm{dis}}$, considered in the symmetric case
$p=1/2$ with $q>q_c$, each realization starts from a fully ordered state. The run is stopped when the total magnetization first reaches or crosses zero, corresponding to the first arrival at the mixed state $c=1/2$ or its immediate finite-jump neighborhood. This convention is appropriate for multispin updates, where a single hyperedge update can cross the mixed state without landing exactly on it.

For the exit probability $E(c_0)$, each realization starts from the prescribed initial fraction $c_0$ and is evolved until it reaches either $c=1$ or $c=0$. The exit probability is estimated as the fraction of runs that reach the all-$+1$ state first.

All observables are averaged over the number of independent runs stated in the corresponding figure captions. In the numerical implementation, a sufficiently large upper cutoff on the total number of elementary updates is imposed for computational safety; for the results reported here, this cutoff does not affect the quoted averages within the displayed accuracy.

\bibliographystyle{apsrev4-2}
\bibliography{apssamp}

\end{document}